\documentclass[11pt,a4paper]{article}
\usepackage{jheppub_kim}
\usepackage{subfigure,amsmath,amssymb ,amsfonts, latexsym}
\usepackage[notcite,color,final]{showkeys}

\newcommand{\be}{\begin{equation}}
\newcommand{\ee}{\end{equation}}
\newcommand{\beqa}{\begin{subequations}\begin{eqnarray}}
\newcommand{\eeqa}{\end{eqnarray}\end{subequations}}

\title{Exact charged black-hole solutions in D-dimensional $f(T)$ gravity:
torsion vs curvature analysis}

\author[a,b]{S. Capozziello}
\author[c,d]{P. A. Gonz\'{a}lez}
\author[e,f,g]{E. N. Saridakis}
\author[h]{ Y. V\'{a}squez}

\affiliation[a]{Dipartimento di Scienze Fisiche, Universit`a di Napoli
\textquotedblleft
Federico II\textquotedblright , Napoli, Italy}

\affiliation[b]{INFN Sez. di Napoli, Compl. Univ. di Monte S. Angelo,
Edificio G, Via
Cinthia, I-80126, Napoli, Italy}

\affiliation[c]{ Escuela de Ingenier\'{\i}a Civil en Obras Civiles.
Facultad de Ciencias F\'{\i}sicas y Matem\'{a}ticas, Universidad Central de
Chile, Avenida Santa
Isabel 1186, Santiago, Chile.}
\affiliation[d]{Universidad Diego Portales, Casilla 298-V, Santiago,
Chile.}

\affiliation[e]{Physics Division, National Technical University of Athens,
15780 Zografou Campus, Athens, Greece}

\affiliation[f]{CASPER, Physics Department, Baylor University,
Waco, TX  76798-7310, USA}

\affiliation[g]{Institut d'Astrophysique de Paris, UMR 7095-CNRS,
Universit\'e Pierre \& Marie Curie, 98bis boulevard Arago, 75014 Paris,
France}

\affiliation[h]{Departamento de Ciencias F\'{\i}sicas, Facultad de
Ingenier\'{i}a, Ciencias
y Administraci\'{o}n, Universidad de La Frontera, Avenida Francisco Salazar
01145, Casilla 54-D, Temuco, Chile.}

\emailAdd{capozzie@na.infn.it}
\emailAdd{pgonzalezm@ucentral.cl}
\emailAdd{Emmanuel$\_$Saridakis@baylor.edu}
\emailAdd{yvasquez@ufro.cl}

\abstract{We extract exact charged black-hole solutions with flat
transverse sections in the framework of D-dimensional Maxwell-$f(T)$
gravity, and we analyze the singularities and horizons based on both
torsion and curvature invariants. Interestingly enough, we find that in
some particular solution subclasses there appear more singularities in the
curvature scalars than in the torsion ones. This difference disappears in
the uncharged case, or in the case where $f(T)$ gravity becomes the usual
linear-in-$T$ teleparallel gravity, that is General Relativity. Curvature
and torsion invariants behave very differently when matter fields are
present, and thus $f(R)$ gravity and $f(T)$ gravity exhibit
different features and cannot be directly re-casted each other.}

\begin{document}

\maketitle

\section{Introduction}

Teleparallel equivalent of General Relativity (TEGR) \cite{ein28,Hayashi79}
is an equivalent formulation of gravity, but, instead of using  curvature
invariants defined by the Levi-Civita connection, the Weitzenb{\"o}ck
connection is adopted. Therefore TEGR exhibits no curvature but only
torsion. The dynamical objects in such a framework are the four linearly
independent {\it vierbeins} and the advantage of this framework is that the
torsion tensor is formed solely by products of first derivatives of these
vierbeins. In such a formulation, as described in \cite{Hayashi79}, the
Lagrangian density, $T$, can then be constructed from this torsion tensor
assuming the invariance under general coordinate transformations, global
Lorentz transformations, and the parity operation, along with demanding the
Lagrangian density to be second order in the torsion tensor. In
\cite{fT,Bengochea:2008gz,Ferraro:2008ey,Linder:2010py} an extension of
the above idea was constructed,  making the Lagrangian density a function
of $T$, similar to the $f(R)$ extension of the Hilbert-Einstein  action.
$f(T)$ gravity has gained a significant attention in the literature, and
proves to exhibit interesting cosmological implications \cite{fT,
Ferraro:2008ey,Bengochea:2008gz,
Linder:2010py,Myrzakulov:2010vz,Chen001,Wu001,
Dent001,Zheng:2010am,Bamba:2010wb,Yang:2010ji,
Wu:2010mn,Bengochea001,Wu:2010xk,Zhang:2011qp,Cai:2011tc,
Chattopadhyay001,Sharif001,Wei001,Boehmer004,Wei005,
Capozziello006,Daouda001,Bamba:2011pz,Geng:2011aj,
Wei:2011yr,Geng:2011ka,Bohmer:2011si,
Atazadeh:2011aa,Jamil:2011mc,Farajollahi:2011af,
Karami:2012fu,Yang:2012hu,Xu:2012jf,Bamba:2012vg,Setare:2012vs,Liu:2012kk,
Iorio:2012cm,
Dong:2012en,Daouda:2012wt,bambaodi,Behboodi:2012ak,Banijamali:2012nx,
Myrzakulov:2012qp,Liu:2012fk,Wanas:2012pu,Rahaman:2012qk,Ghosh:2012pg,
Rodrigues:2012qu}.

Such an approach  can be framed within the class of new gravity theories
aimed to extend General Relativity in order to solve its shortcomings at
Infra-Red and Ultra-Violet scales \cite{review}. Clearly, in extending the
geometry sector, one of the goals is to solve the puzzle of dark energy and
dark matter that, up to now, seems to have no counterpart at fundamental
level. In other words,  both $f(T)$ gravity and $f(R)$ gravity could be
reliable approaches to address the problems of missing matter and
accelerated expansion without asking for new material ingredients that have
not been detected yet by the experiments \cite{francaviglia}.
   
In this work we investigate D-dimensional $f(T)$ gravity, considering
additionally the electromagnetic sector. Exact black-hole solutions with
flat transverse section (Banados-Teitelboim-Zanelli (BTZ)-like solutions 
\cite{BTZ}) can be derived for a given range of parameter space. Then we
analyze the singularities of these solutions based on the torsion scalar
and the curvature scalar, pointing out differences with respect to $f(R)$
gravity. It is important to stress that searching for exact solutions is a
fundamental step to set a new field theory. Exact solutions allow a full
control of the systems and can contribute to the well-formulation and
well-position of the Cauchy problem (for a discussion on this point see
\cite{libro}).

The paper is organized  as follows. In Sec. \ref{TEGR}, we present a
brief review of TEGR in four dimensions, as well as of its $f(T)$
extension. In Sec. \ref{TelD} the D-dimensional teleparallel gravity is
formulated and  the analysis is extended  to D-dimensional Maxwell-$f(T)$
gravity. In Sec. \ref{exactsol} we derive  exact charged static solutions
and  Sec. \ref{singularities} is devoted to the  investigation of 
singularities and horizons. Finally, in Sec. \ref{conclusions} we discuss
some physical implications of the results.

\section{Teleparallel Equivalent to General Relativity and its
$f(T)$ extension}
\label{TEGR}

In this section we briefly review Teleparallel Equivalent to General
Relativity (TEGR) in four dimensions and its $f(T)$
extension. Throughout the manuscript we use the following notation:
Greek indices $\mu, \nu,$... run over all space-time coordinates  0, 1, 2,
3; lower case Latin indices (from the middle of the alphabet) $i, j, ...$
run over spatial coordinates 1, 2, 3; capital Latin indices $A, B, $... run
over the tangent space-time 0, 1, 2, 3, and lower case Latin indices (from
the beginning of the alphabet) $a,b, $... run over the tangent space
spatial coordinates 1, 2, 3.

As we mentioned above, the dynamical variable of teleparallel gravity is
the vierbein field ${\mathbf{e}_A(x^\mu)}$, which forms an orthonormal
basis for the tangent space at each point $x^\mu$ of the manifold, that is
$\mathbf{e} _A\cdot\mathbf{e}_B=\eta_{AB}$, with
$\eta_{AB}=diag (1,-1,-1,-1)$. Moreover, the vector $\mathbf{e}_A$ can
be expressed in terms of its components $e_A^\mu$ in a coordinate basis,
namely $\mathbf{e}_A=e^\mu_A\partial_\mu $. In such a formulation the
metric tensor is acquired from the dual vierbein as 
\begin{equation}  \label{metrdef}
g_{\mu\nu}(x)=\eta_{AB}\, e^A_\mu (x)\, e^B_\nu (x)~.
\end{equation}
Although in General Relativity one uses the torsionless
Levi-Civita connection, in the present construction one uses the
curvatureless Weitzenb\"{o}ck connection \cite{Weitzenb23}, whose torsion
tensor reads 
\begin{equation}  \label{torsion2}
{T}^\lambda_{\:\mu\nu}=\overset{\mathbf{w}}{\Gamma}^\lambda_{ \nu\mu}-
\overset{\mathbf{w}}{\Gamma}^\lambda_{\mu\nu} =e^\lambda_A\:(\partial_\mu
e^A_\nu-\partial_\nu e^A_\mu)~.
\end{equation}
Moreover, the contorsion tensor, which gives the difference between
Weitzenb\"{o}ck and Levi-Civita connections, is given by 
$K^{\mu\nu}_{\:\:\:\:\rho}=-\frac{1}{2}\left(T^{\mu\nu}_{ \:\:\:\:\rho}
-T^{\nu\mu}_{\:\:\:\:\rho}-T_{\rho}^{\:\:\:\:\mu\nu}\right)$, while it
proves convenient to define
$S_\rho^{\:\:\:\mu\nu}=\frac{1}{2}\left(K^{\mu\nu}_{\:\:\:\:\rho}
+\delta^\mu_\rho \:T^{\alpha\nu}_{\:\:\:\:\alpha}-\delta^\nu_\rho\:
T^{\alpha\mu}_{\:\:\:\:\alpha}\right)$.
For a detailed exposition of torsion tensor properties see
\cite{annalen}.

In conclusion, in the present formulation the torsion tensor ${T}
^\lambda_{\:\mu\nu} $ includes all the information concerning the
gravitational field. Using the above definitions one can construct the
simplest form of the ``teleparallel'' Lagrangian, which is the torsion
scalar, that is \cite{Maluf:1994ji,Arcos:2005ec}  
\begin{equation}  \label{telelag}
\mathcal{L}=T\equiv\frac{1}{4}
T^{\rho \mu \nu}T_{\rho \mu \nu}+\frac{1}{2}T^{\rho \mu \nu}T_{\nu \mu
\rho}-T_{\rho \mu}^{\ \ \rho}T_{\ \ \ \nu}^{\nu \mu}~.
\end{equation}
  Thus, the simplest action of teleparallel gravity reads: 
\begin{eqnarray}  \label{action0}
S = \frac{1}{2 \kappa}\int d^4x e \left(T+\mathcal{L}_{m}\right)~,
\end{eqnarray}
where $\kappa =8 \pi G$, $e = \text{det}(e_{\mu}^A) = \sqrt{-g}$ and
$\mathcal{L}_{m}$ accounts for the matter Lagrangian. It is worth
noticing here that the Ricci scalar $R$ and the torsion scalar $T$ differ
only by a total derivative of the torsion tensor, namely 
\cite{Weinberg:2008}:
\begin{equation}  \label{RTrelation}
R=-T-2\nabla^\mu\left(T^\nu_{\ \mu\nu}\right).
\end{equation}

Varying the action (\ref{action0}) with respect to the vierbein we
obtain the field equations 
\begin{eqnarray}  \label{eom}
e^{-1}\partial_{\mu}(ee_A^{\rho}S_{\rho}{}^{\mu\nu})
-e_{A}^{\lambda}T^{\rho}{}_{\mu\lambda}S_{\rho}{}^{\nu\mu} -\frac{1}{4}
e_{A}^{\nu }T = 4\pi Ge_{A}^{\rho}\overset{\mathbf{em}}{T}_{\rho}{}^{\nu}~,
\end{eqnarray}
where the tensor
$\overset{\mathbf{em}}{T%
}_{\rho}{}^{\nu}$ on the right-hand side is the usual energy-momentum
tensor of matter fields. These equations coincide with those of
General Relativity for every geometry choice, and this is the why the
theory is named ``Teleparallel Equivalent to General Relativity''.

One can generalize the above formulation considering arbitrary functions of
the torsion scalar $f(T)$ in the gravitational action 
\cite{fT, Ferraro:2008ey,Bengochea:2008gz,Linder:2010py}, although the
Lorentz invariance of the linear theory seems to be spoiled
\cite{fTLorinv0,fTLorinv2}. Thus, the action becomes
\begin{equation}
\label{actionfT4D0}
S = \frac{1}{2 \kappa}\int d^4x e
\left[T+f(T)+\mathcal{L}_{m}\right].
\end{equation}
Notice the difference in the various conventions in 4D-$f(T)$ literature,
since some authors replace $T$ by $f(T)$, while the majority replace $T$ by
$T+f(T)$. In this work we follow the second convention, that is
teleparallel gravity  is acquired by setting $f(T)=0$. Finally, variation
of the action (\ref{actionfT4D0}) with respect to the vierbein gives the
field equations 
\begin{equation}\label{eomfT4D}
e^{-1}\partial_{\mu}(e
e_A^{\rho}S_{\rho}{}^{\mu\nu})\left(1+\frac{df}{dT}\right)
-e_{A}^{\lambda}T^{\rho}{}_{\mu\lambda}S_{\rho}{}^{\nu\mu} +
e_A^{\rho}S_{\rho}{}^{\mu\nu}\partial_{\mu}({T})\frac{d^2f}{dT^2}-\frac{1}{
4 } e_ { A } ^ {
\nu
}[T+f({T})]
= 4\pi Ge_{A}^{\rho}\overset {\mathbf{em}}T_{\rho}{}^{\nu}.
\end{equation}

\section{D-dimensional Teleparallel Gravity  and its Maxwell-$f(T)$
extension}
\label{TelD}

In this section we present teleparallel gravity in D-dimensions  and its
Maxwell-$f(T)$ extension and we explore its properties. It proves more
convenient to use differential forms, where the torsion 2-form $T^{a}$  is
simply $T^{a}=de^{a}$.

We start with the gravitational teleparallel action with the most general 
quadratic form in the torsion tensor. Under the assumption
of zero spin-connection it is given by \cite{Muench:1998ay,Itin:1999wi} 
\begin{equation}
S=\frac{1}{2\kappa }\int \left( \rho _{0}\mathcal{L}_{0}+\rho
_{1}\mathcal{L}
_{1}+\rho _{2}\mathcal{L}_{2}+\rho _{3}\mathcal{L}_{3}+\rho _{4}\mathcal{L}
_{4}\right) ~,  \label{action2}
\end{equation}
where $\rho _{i}$ are dimensionless parameters and 
\begin{equation}
\mathcal{L}_{0}=\frac{1}{4}e^{a}\wedge \star e_{a}~,\quad \mathcal{L}
_{1}=de^{a}\wedge \star de_{a}~,\quad \mathcal{L}_{2}=(de_{a}\wedge \star
e^{a})\wedge \star (de_{b}\wedge e^{b})~,  \notag
\end{equation}
\begin{equation}
\mathcal{L}_{3}=(de^{a}\wedge e^{b})\wedge \star (de_{a}\wedge
e_{b})~,\quad 
\mathcal{L}_{4}=(de_{a}\wedge \star e^{b})\wedge \star (de_{b}\wedge
e^{a})~,
\end{equation}
with $\star $ standing for the Hodge dual operator and $\wedge $ for the
usual
wedge product. The coupling constant $\rho _{0}=-\frac{8}{3}\Lambda $
accounts for the cosmological constant term, and furthermore, since
$\mathcal{L}_{3}$ can be
completely expressed in terms of $\mathcal{L}_{1}$, in the following we set
$\rho _{3}=0$  \cite{Muench:1998ay}. Lastly, we mention that in the above
expression $\kappa $ is the D-dimensional gravitational constant, while
the vierbeins and the metric are now D-dimensional. Therefore, in the
following, all the conventions  adopted in  Sec. \ref{TEGR} extend in D
dimensions.

Action (\ref{action2}) can be written more conveniently as 
\begin{equation}
S=\frac{1}{2\kappa }\int \left( T-2\Lambda \right) \star
1~=\frac{1}{2\kappa }\int d^{D}x\,e\left( T-2\Lambda \right),
\label{actiontele0}
\end{equation}
where $\star 1=(-1)^{D-1}e^{0}\wedge e^{1}\wedge e^{2}......\wedge e^{D-1}$, and the
torsion scalar $T$ is given by 
\begin{equation}
T=(-1)^{D-1}\star \left[ \rho _{1}(de^{a}\wedge \star de_{a})+\rho
_{2}(de_{a}\wedge
e^{a})\wedge \star (de_{b}\wedge e^{b})+\rho _{4}(de_{a}\wedge e^{b})\wedge
\star (de_{b}\wedge e^{a})\right].  \label{scalartorsion}
\end{equation}
Expanding this expression in its components we acquire
\begin{equation}
T=\frac{1}{2}\left(\rho _{1}+\rho _{2}+\rho _{4}\right)T^{abc}T_{abc}+\rho
_{2}T^{abc}T_{bca}-\rho _{4}T_{a}^{ac}T_{bc}^{b}~, 
\label{scalartorsionrho}
\end{equation}
thus we straightforwardly see that for $\rho _{1}=0$, $\rho _{2}=-%
\frac{1}{2}$ and $\rho _{4}=1$ it coincides with (\ref{telelag}) in D
dimensions, namely 
\begin{equation}
T=\frac{1}{4}T^{abc}T_{abc}-\frac{1}{2}
T^{abc}T_{bca}-T_{a}{}^{ac}T^{b}{}_{bc}~.  \label{actiontel3D}
\end{equation}

Now, we will extend the above discussion considering arbitrary functions of
the torsion scalar $f(T)$ in the D-dimensional gravitational action.  Thus,
we
consider
an action of the form 
\begin{equation}
S=\frac{1}{2\kappa }\int d^{D}xe\left[ T+f(T)-2\Lambda \right] ~,
\label{actionfT3D0}
\end{equation}%
with the torsion scalar $T$ given by (\ref{scalartorsionrho}), that is we
keep the general coefficients $\rho _{i}$. In differential forms the above
action can be written as
\begin{equation}
S=\frac{1}{2\kappa }\int \left\{ \left[ f(T)+T-2\Lambda \right] \star
1\right\} ~,  \label{accionfTdif}
\end{equation}
where now $T$ is given by (\ref{scalartorsion}). Finally, note that
teleparallel
D-dimensional gravity discussed above is obtained by setting 
$f(T)=0$.

Lastly, we extend the discussion incorporating additionally the
electromagnetic
sector. In particular, we extend the total action to 
\begin{equation}
S=\frac{1}{2\kappa }\int \left\{ \left[ f(T)+T-2\Lambda \right] \star
1\right\} +\int \mathcal{L}_{F}~,  \label{accionfTMdif}
\end{equation}
where 
\begin{equation}
\mathcal{L}_{F}=-\frac{1}{2}F\wedge ^{\star }F
  \label{Mlagrangian}
\end{equation}
is the Maxwell Lagrangian, while $F=dA$, with $A\equiv A_{\mu
}dx^{\mu }$, is the electromagnetic potential 1-form. The action variation
leads to the following field equations: 
\begin{eqnarray}
\delta \mathcal{L} &=&\delta e^{a}\wedge \left\{ \left( 1+\frac{df}{dT}
\right) \left\{ \rho _{1}\left[ 2d\star de_{a}+i_{a}(de^{b}\wedge \star
de_{b})-2i_{a}(de^{b})\wedge \star de_{b}\right] \right. \right.  \notag
\label{fieldeq} \\
&&\ \ \ \ \ \ \ \ +\rho _{2}\left\{ -2e_{a}\wedge d\star (de^{b}\wedge
e_{b})+2de_{a}\wedge \star (de^{b}\wedge e_{b})+i_{a}\left[ de^{c}\wedge
e_{c}\wedge \star (de^{b}\wedge e_{b})\right] \right.  \notag \\
&&\left. \ \ \ \ \ \ \ \ \ \ \ \ \ \ \ \,-2i_{a}(de^{b})\wedge e_{b}\wedge
\star (de^{c}\wedge e_{c})\right\}  \notag \\
&&\ \ \ \ \ \ \ \ +\rho _{4}\left\{ -2e_{b}\wedge d\star (e_{a}\wedge
de^{b})+2de_{b}\wedge \star (e_{a}\wedge de^{b})\right.  \notag \\
&&\left. \left. \ \ \ \ \ \ \ \ \ \ \ \ \ \ \ \,+i_{a}\left[ e_{c}\wedge
de^{b}\wedge \star (de^{c}\wedge e_{b})\right] -2i_{a}(de^{b})\wedge
e_{c}\wedge \star (de^{c}\wedge e_{b})\right\} \right\}  \notag \\
&&\ \ \ \ \ \ \ \ +2\frac{d^{2}f}{dT^{2}}dT\left[ \rho _{1}\star
de_{a}+\rho
_{2}e_{a}\wedge \star (de_{b}\wedge e^{b})+\rho _{4}e_{b}\wedge \star
(de^{b}\wedge e_{a})\right]  \notag \\
&&\left. \ \ \ \ \ \ \ \ +\left[ f(T)-T\frac{df}{dT}\right] \wedge \star
e_{a}-2\Lambda \star e_{a}-\frac{1}{2}\left[ F\wedge i_{a}\left( \star
F\right) -i_{a}\left( F\right) \wedge \star F\right] \right\}  \notag \\
&&\ \ \ \ \ \ \ \ \ +\delta A\left( d^{\star }F\right) =0~,
\label{maxwellfieldeq}
\end{eqnarray}
where $i_{a}$ is the interior product. Although one could investigate
solution subclasses with general coupling parameters $\rho _{i}$, in the
following,  for the sake of  simplicity,  we restrict to the standard  case $\rho _{1}=0$, $\rho
_{2}=-1/2$ and $\rho _{4}=1$ of (\ref{actiontel3D}).

\section{Exact charged solutions}
\label{exactsol}

Let us  now investigate the charged solutions of the theory. In order to
extract the static solutions we consider
the metric form 
\begin{equation}\label{metric}
ds^{2}=F\left( r\right) ^{2}dt^{2}-\frac{1}{G\left( r\right) ^{2}}
dr^{2}-r^{2}\sum_{i=1}^{i=D-2}dx_{i}^{2}~,
\end{equation}
which arises from the vierbein 
diagonal ansatz 
\begin{equation}\label{diagonal}
e^{0}=F\left( r\right) dt~,\text{ \ }e^{1}=\frac{1}{G\left( r\right) }dr~,
\text{ \ }e^{2}=rdx_{1}~,\text{ \ \ }e^{3}=rdx_{2}~, \text{ \  }\ldots.
\end{equation}
Let us make an important comment here concerning the vierbein choice that
corresponds to the metric (\ref{metric}). In the case of
linear-in-$T$  gravity the above  simple, diagonal  relation 
between the metric (\ref{metric}) and the vierbeins (\ref{diagonal}) is
always allowed. On the contrary, in the extension of $f(T)$ gravity, in
general, one could have a more complicated relation connecting the vierbein
with the metric, with the vierbein being non-diagonal even for a diagonal
metric
\cite{Ferraro001,Ferraro002,Tamanini:2012hg,Daouda:2012nj}.
However, in the cosmological investigations of
$f(T)$ gravity \cite{fT,Ferraro:2008ey,Bengochea:2008gz,
Linder:2010py,Myrzakulov:2010vz,Chen001,Wu001,
Dent001,Zheng:2010am,Bamba:2010wb,Yang:2010ji,
Wu:2010mn,Bengochea001,Wu:2010xk,Zhang:2011qp,Cai:2011tc,
Chattopadhyay001,Sharif001,Wei001,Boehmer004,Wei005,
Capozziello006,Daouda001,Bamba:2011pz,Geng:2011aj,
Wei:2011yr,Geng:2011ka,Bohmer:2011si,
Atazadeh:2011aa,Jamil:2011mc,Farajollahi:2011af,
Karami:2012fu,Yang:2012hu,Xu:2012jf,Bamba:2012vg,Setare:2012vs,Liu:2012kk,
Iorio:2012cm,
Dong:2012en,Daouda:2012wt,bambaodi,Behboodi:2012ak,Banijamali:2012nx,
Myrzakulov:2012qp,Liu:2012fk,Wanas:2012pu,Rahaman:2012qk,Ghosh:2012pg,
Rodrigues:2012qu},
as well as in
its black-hole solutions
\cite{Wang:2011xf,Miao003,Wei:2011aa,Ferraro:2011ks,Gonzalez:2011dr},
the authors still use the diagonal relation between the vierbeins and the
metric, as a first approach to reveal the features of the theory.
Thus, in the present investigation we also impose the diagonal relation
between the vierbeins and the metric, as a first approach on the subject
and
in order to reveal the main features of the solution structure. However,
we are aware that a detailed study of the general vierbein choice (and its
relation to extra degrees of freedom) is a necessary step for a deeper
understanding of $f(T)$-gravity foundations.

Concerning the electric sector of the electromagnetic 2-form we assume
\begin{equation}
F=dA=E_{r}\left( r\right) e^{1}\wedge e^{0}+\sum_{i=1}^{i=D-2}E_{i}( r
)e^{0}\wedge e^{i+1}~,
\end{equation}
where $E_{r}$ is the radial electric field, neglecting for the moment the
magnetic part.
Thus, inserting the above ansatzes
in the field equations (\ref{maxwellfieldeq}), we finally obtain 
\begin{equation}
\left( 1+\frac{df}{dT}\right) T-\left[ f\left( T\right) -T\frac{df}{dT}%
\right] +2\Lambda
+\frac{1}{2}E_{r}^{2}-\frac{1}{2}\sum_{i=1}^{i=D-2}E_{i}^{2}=0~,   
\label{fieldequation1}
\end{equation}
\begin{equation}
\left( 1+\frac{df}{dT}\right) \left[ -\frac{G\left( r\right) G^{\prime
}\left( r\right) }{r}+\frac{F^{\prime }\left( r\right) G\left( r\right)
^{2}%
}{rF\left( r\right) }\right] -\frac{d^{2}f}{dT^{2}}T^{\prime }\left(
r\right) \frac{G\left( r\right)
^{2}}{r}-\frac{1}{2}\sum_{i=1}^{i=D-2}E_{i}^{2}=0~,  
\label{fieldequation2}
\end{equation}
\begin{eqnarray}
\nonumber\left( 1+\frac{df}{dT}\right) \left[ -\frac{F^{\prime \prime
}\left(
r\right) G\left( r\right) ^{2}}{F\left( r\right) }-\frac{F^{\prime }\left(
r\right) G^{\prime }\left( r\right) G\left( r\right) }{F\left( r\right) }+
\frac{F^{\prime }\left( r\right) G\left( r\right) ^{2}}{rF\left(
r\right)}\right]&&\\ \nonumber
&&\\-\nonumber\left( 1+\frac{df}{dT}\right) \left[\left( D-3\right)
\frac{G^{\prime }\left( r\right) G\left( r\right) }{r}
-\left( D-3\right) \frac{G\left( r\right) ^{2}}{r^{2}}\right]&&\\ 
-\frac{d^{2}f}{dT^{2}}T^{\prime }\left( r\right) \left[ \frac{F^{\prime
}\left( r\right) G\left( r\right) ^{2}}{F\left( r\right) }+\left(
D-3\right) 
\frac{G\left( r\right) ^{2}}{r}\right] +\frac{1}{2}E_{r}^{2}-\frac{1}{2}%
E_{1}^{2}=0~,\label{fieldequation3} 
\end{eqnarray}
\begin{equation}
E_{r}E_{j}=0\text{ \ \ }j=1,\ldots ,D-2~, 
\label{electric1}
\end{equation}
\begin{equation}
E_{i}E_{j}=0\text{ \ \ }i,j=1,\ldots ,D-2~(i\ne j)~,  
\label{electric2}
\end{equation}
where
\begin{equation}
T\left( r\right) =2\left( D-2\right) \frac{F^{\prime }\left( r\right)
G\left( r\right) ^{2}}{rF\left( r\right) }+\left( D-2\right) \left(
D-3\right) \frac{G\left( r\right) ^{2}}{r^{2}}\text{ }.
\label{scalartorsion1}
\end{equation}
The remaining field equations are equivalent to equation
(\ref{fieldequation3}), that is the $a=j$ equation is similar to
(\ref{fieldequation3}), but with $-\frac{1}{2}E_{1}^{2}$ replaced by
$-\frac{1}{2}E_{j-1}^{2}$ .

A first observation is that from (\ref{fieldequation1}) we deduce that $T$
has, in general, an $r$-dependence, which disappears for a zero
electric charge. Such a behavior reveals the new features that are brought
in by the richer structure of the addition of the electromagnetic sector.
Moreover, form (\ref{electric1}) and (\ref{electric2}), we deduce that we
cannot have simultaneously two non-zero electric field components. This
result
is similar to the known no-go theorem of 3D GR-like gravity
\cite{Cataldo:2002fh,Blagojevic00}, which states that configurations with
two non-vanishing components of the Maxwell field are dynamically not
allowed. However, it is not valid anymore if we add the magnetic sector, as
we will see in subsection \ref{magnfield} (it holds only for D=3).
Therefore, in the following we investigate the cases of radial electric
field, of non-radial electric field, and of magnetic and radial
electric field, separately.

\subsection{Radial electric field}

We first consider the case where there exists only radial electric field.
Thus, the Maxwell equations give
\begin{equation}
E_{r}=\frac{Q}{r^{D-2}}~,
\end{equation}
where $Q$ is an integration constant which, as usual, coincides with the
electric charge of the black hole. Now, integrating equation
(\ref{fieldequation2}) we find the very simple and helpful result
\begin{equation}
F\left( r\right) =G\left( r\right) \left( 1+\frac{df}{dT}\right)~.
\label{expression}
\end{equation}
Using equations (\ref{expression}) and (\ref{scalartorsion1}) we obtain
\begin{equation}
\frac{dG\left( r\right) ^{2}}{dr}+\left[ 2\frac{d}{dr}\ln \left(
1+\frac{df}{dT}\right) +\frac{\left( D-3\right) }{r}\right] G\left(
r\right) ^{2}-\frac{rT\left( r\right) }{\left( D-2\right) }=0~,
\end{equation}
whose solution is
\begin{equation}\label{G}
G\left( r\right) ^{2}=\frac{1}{\left( 1+\frac{df}{dT}\right) ^{2}r^{D-3}}
\left[ \frac{1}{\left( D-2\right) }\int \left( 1+\frac{df}{dT}\right)
^{2}r^{D-2}T\left( r\right) dr+Const\right]~,
\end{equation}
and using equation (\ref{expression}) we get
\begin{equation}
\label{Fsol}
F\left( r\right) ^{2}=\frac{1}{r^{D-3}}\left[ \frac{1}{\left( D-2\right) }
\int \left( 1+\frac{df}{dT}\right) ^{2}r^{D-2}T\left( r\right)
dr+Const\right]~,
\end{equation}
where $Const$ is an integration constant related to the mass of the
spherical object.

In order to proceed, and similar to \cite{Gonzalez:2011dr}, we will
consider
Ultraviolet (UV) corrections of $f(T)$ gravity. In particular, we examine
the modifications on the solutions caused by UV
modifications of  D-dimensional gravity and we consider a representative
ansatz
of the form $f(T)=\alpha T^{2}$. This is the first order correction in
every realistic $f(T)$ gravity, in which we expect $f(T)\ll T$
\cite{Wu:2010mn,Iorio:2012cm}, since $T$ (like $R$) is small in
$\kappa^2$-units. Thus, for $\alpha \neq 0$, equation
(\ref{fieldequation1}) leads to 
\begin{equation}
\label{TUV}
T\left( r\right) =\frac{-1\pm \sqrt{1-24\alpha \Lambda -6\alpha
Q^{2}r^{4-2D}
}}{6\alpha}~,
\end{equation}
with the upper and lower signs corresponding to the positive and negative
branch solutions respectively
(note that if $\alpha=0$ then (\ref{fieldequation1}) becomes linear having
only one solution, which is given by the $\alpha\rightarrow0$ limit of the
positive branch of (\ref{TUV}), namely $T(r)=-Q^2 r^{4-2D}/2-2\Lambda$, in
which case teleparallel gravity is restored). Thus,
\begin{equation}
\label{dfdTUV}
1+\frac{df}{dT}=\frac{2}{3}\pm \frac{1}{3}\sqrt{1-24\alpha \Lambda -6\alpha
Q^{2}r^{4-2D}}\text{ },
\end{equation}
and therefore performing the integration that appears in
(\ref{G}) and (\ref{Fsol}), we obtain
\begin{eqnarray}
\nonumber \int \left(1+\frac{df}{dT}\right) ^{2}r^{D-2}T\left( r\right)
dr=\frac{1}{54\alpha} 
\left[-\frac{18\alpha Q^{2}r^{3-D}}{3-D}-\frac{\left( 1+72\alpha \Lambda
\right) r^{D-1}}{D-1}\right]&&\\
\nonumber &&\\\nonumber  \pm \frac{\sqrt{r^{4D}\left( 1-24\alpha \Lambda
-6\alpha
Q^{2}r^{4-2D}\right) }}{54\alpha}\left[ \frac{6\alpha
Q^{2}r^{3-3D}}{2D-5}-\frac{\left(
-1+24\alpha \Lambda \right) r^{-1-D}}{D-1}\right]&&\\
\nonumber&&\\ \mp \frac{\left( D-2\right) ^{2}\left( -1+24\alpha \Lambda
\right) Q^{2}r^{3+D}
\sqrt{1+\frac{6\alpha Q^{2}r^{4-2D}}{-1+24\alpha \Lambda }}\ _2F_1\left(
\frac{D-3}{2\left( D-2\right) },\frac{1}{2},\frac{3D-7}{2\left( D-2\right)
};\frac{6\alpha Q^{2}r^{4-2D}}{1-24\alpha \Lambda }\right)}{ 3\left(
D-3\right) \left( 2D-5\right) \left( D-1\right) \sqrt{r^{4D}\left(
1-24\alpha \Lambda -6\alpha Q^{2}r^{4-2D}\right)}},\ \ \ \ \,
\label{integral}
\end{eqnarray}
where 
$ \
_2F_1(a,b,c;x)$ is the hypergeometric function. We mention that the last
argument of this function, namely $\left( 6\alpha Q^{2}r^{4-2D}\right)
/\left(1-24\alpha \Lambda \right)$, must be negative, while from
(\ref{TUV}) it is required that  $1-24\alpha \Lambda
-6\alpha Q^{2}r^{4-2D}$ must be positive, therefore we deduce that
$\alpha$ should be negative.

In summary, inserting the integral (\ref{integral}), along with
(\ref{dfdTUV}), in (\ref{G}) and (\ref{Fsol}), we find that the black-hole
solution is:
\begin{eqnarray}\label{G2}
&&G\left( r\right) ^{2}=\frac{1}{\left( \frac{2}{3}\pm
\frac{1}{3}\sqrt{1-24\alpha \Lambda -6\alpha
Q^{2}r^{4-2D}}\right) ^{2}r^{D-3}}
\Big\{ \frac{1}{\left( D-2\right) }\nonumber\\
&&\Big\{ \frac{1}{54\alpha} 
\Big[-\frac{18\alpha Q^{2}r^{3-D}}{3-D}-\frac{\left( 1+72\alpha \Lambda
\right) r^{D-1}}{D-1}\Big]\nonumber\\
 &&  \pm \frac{\sqrt{r^{4D}\left( 1-24\alpha \Lambda
-6\alpha
Q^{2}r^{4-2D}\right) }}{54\alpha}\Big[ \frac{6\alpha
Q^{2}r^{3-3D}}{2D-5}-\frac{\left(
-1+24\alpha \Lambda \right) r^{-1-D}}{D-1}\Big]\nonumber\\
 && \mp \frac{\left( D-2\right) ^{2}\left( -1+24\alpha \Lambda
\right) Q^{2}r^{3+D}
\sqrt{1+\frac{6\alpha Q^{2}r^{4-2D}}{-1+24\alpha \Lambda }}\ _2F_1\left(
\frac{D-3}{2\left( D-2\right) },\frac{1}{2},\frac{3D-7}{2\left( D-2\right)
};\frac{6\alpha Q^{2}r^{4-2D}}{1-24\alpha \Lambda }\right)}{ 3\left(
D-3\right) \left( 2D-5\right) \left( D-1\right) \sqrt{r^{4D}\left(
1-24\alpha \Lambda -6\alpha Q^{2}r^{4-2D}\right)}}\Big\}\nonumber\\
&&+Const\Big\} \ \ \
\ \  \ 
\end{eqnarray}
and  
\begin{eqnarray}
\label{Fsol2}
&&F\left( r\right) ^{2}=\frac{1}{r^{D-3}}
\Big\{ \frac{1}{\left( D-2\right) }\nonumber\\
&&\Big\{ \frac{1}{54\alpha} 
\Big[-\frac{18\alpha Q^{2}r^{3-D}}{3-D}-\frac{\left( 1+72\alpha \Lambda
\right) r^{D-1}}{D-1}\Big]\nonumber\\
 &&  \pm \frac{\sqrt{r^{4D}\left( 1-24\alpha \Lambda
-6\alpha
Q^{2}r^{4-2D}\right) }}{54\alpha}\Big[\frac{6\alpha
Q^{2}r^{3-3D}}{2D-5}-\frac{\left(
-1+24\alpha \Lambda \right) r^{-1-D}}{D-1}\Big]\nonumber\\
 && \mp \frac{\left( D-2\right) ^{2}\left( -1+24\alpha \Lambda
\right) Q^{2}r^{3+D}
\sqrt{1+\frac{6\alpha Q^{2}r^{4-2D}}{-1+24\alpha \Lambda }}\ _2F_1\left(
\frac{D-3}{2\left( D-2\right) },\frac{1}{2},\frac{3D-7}{2\left( D-2\right)
};\frac{6\alpha Q^{2}r^{4-2D}}{1-24\alpha \Lambda }\right)}{ 3\left(
D-3\right) \left( 2D-5\right) \left( D-1\right) \sqrt{r^{4D}\left(
1-24\alpha \Lambda -6\alpha Q^{2}r^{4-2D}\right)}}\Big\}\nonumber\\
 &&+Const\Big\}.\ \ \
\ \  \ 
\end{eqnarray}

The special point in the parameter space $\Lambda =1/(24\alpha )$ needs to
be analyzed separately, since in this point we obtain the solution
\begin{equation}\label{int}
\int \left( 1+\frac{df}{dT}\right) ^{2}r^{D-2}T\left( r\right)
dr=-\frac{r^{-1-5D}}{27\alpha }\left[ \frac{2r^{6D}}{D-1}-\frac{9\alpha
Q^{2}r^{4+4D}}{D-3}\pm \frac{3\sqrt{6}\left( -\alpha Q^{2}r^{4+2D}\right)
^{3/2}}{2D-5}
\right],
\end{equation}
and thus 
\begin{eqnarray}\label{G2spec}
&&G\left( r\right) ^{2}=\frac{1}{\left( \frac{2}{3}\pm
\frac{1}{3}\sqrt{1-24\alpha \Lambda -6\alpha
Q^{2}r^{4-2D}}\right) ^{2}r^{D-3}}
\Big\{ \frac{1}{\left( D-2\right) }\nonumber\\
&&\Big\{ -\frac{r^{-1-5D}}{27\alpha }\Big[
\frac{2r^{6D}}{D-1}-\frac{9\alpha
Q^{2}r^{4+4D}}{D-3}\pm \frac{3\sqrt{6}\left( -\alpha Q^{2}r^{4+2D}\right)
^{3/2}}{2D-5}
\Big]\Big\}+Const\Big\},\ \ \
\ \  \ 
\end{eqnarray}
and  
\begin{eqnarray}
\label{Fsol2spec}
&&F\left( r\right) ^{2}=\frac{1}{r^{D-3}}
\Big\{ \frac{1}{\left( D-2\right) }\nonumber\\
&&\Big\{ -\frac{r^{-1-5D}}{27\alpha }\Big[
\frac{2r^{6D}}{D-1}-\frac{9\alpha
Q^{2}r^{4+4D}}{D-3}\pm \frac{3\sqrt{6}\left( -\alpha Q^{2}r^{4+2D}\right)
^{3/2}}{2D-5}
\Big]\Big\}+Const\Big\}.\ \ \
\ \  \ 
\end{eqnarray}

Finally, the case $D=3$ has to be analyzed separately. Taking properly
the limit $D=3$ of the above expressions we obtain the solutions
extracted in \cite{Gonzalez:2011dr} for 3D Maxwell-$f(T)$ gravity. Lastly,
one can straightforwardly check that in the limit $\alpha\rightarrow0$ (of
the positive branch since in this case the negative branch disappears) one
re-obtains the usual charged General Relativity solutions.

\subsection{Zero radial field}

Let us for the moment assume that we have zero radial field. In this case
equation (\ref{electric2}) implies that we can have at most one non-zero
component of the electric field along the non-radial (transversal)
directions. However,
as we mentioned below equation (\ref{scalartorsion1}), for $D>3$ the
remaining field equations are similar to equation (\ref{fieldequation3})
but with $-\frac{1}{2}E_{1}^{2}$ replaced by $-\frac{1}{2}
E_{j-1}^{2}$, therefore subtracting these equations we acquire the
conditions $E_{i}^{2}=E_{j}^{2}$, with $i$ and $j$ running from $1$ to
$D-2$. These conditions, along with equation (\ref{electric2}), yield
$E_{i}=0$ ($i=1,...,D-2$) for $D>3$, that is the electric field
is completely zero. The only cases where zero radial electric field does
not lead to a disappearance of the total electric field is for $D=3$
(where a non-zero azimuthal electric field is possible) which was analyzed
in detail in \cite{Gonzalez:2011dr}, or if we consider simultaneously
non-zero non-radial electric field with magnetic field, case which lies
beyond the scope of the present investigation.

\subsection{Magnetic field and radial electric field}
\label{magnfield}

For completeness we also examine the case where magnetic field is present.
While in $D=3$ we deduce that electric field must be absent
\cite{Gonzalez:2011dr}, for $D>3$ one can simultaneously have non-zero
magnetic and electric fields. As an explicit example we consider an
electromagnetic strength 2-form in four dimensions given by
\begin{equation}
F=E_{r}\left( r\right) e^{1}\wedge e^{0}+B_{23}\left( r\right) e^{2}\wedge
e^{3}~,
\end{equation}
that is we consider a radial electric field $E_{r}$ and a magnetic
field $B_{23}$ both depending on the radial coordinate $r$ only. From the
Maxwell equations in four dimensions for the electric
field we immediately obtain 
\begin{equation}
E_{r}\left( r\right) =\frac{Q}{r^{2}}~,
\end{equation}
while incorporating the equations of motion analogous to
(\ref{fieldequation1})-(\ref{electric2}) we can see that a
solution is obtained by
\begin{equation}
B_{23}\left( r\right) =\frac{P}{r^{2}},
\end{equation}
leading to the metric coefficients (\ref{Fsol2}) and (\ref{int}) with
$Q^{2}+P^{2}$ in place of $Q^{2}$ (and for $D=4$).

\section{Singularities and horizons}
\label{singularities}

Let us now investigate the singularities and the horizons of the above
solutions. The first step is to find at which $r$ do the functions $G\left(
r\right) ^{2}$ and $F\left( r\right) ^{2}$ become zero or infinity.
However, since these singularities may correspond to  coordinate
singularities, the usual procedure is to investigate various invariants,
since if these invariants diverge at one point they will do that
independently of the specific coordinate basis, and thus the corresponding
point is a physical singularity (note that the opposite is not true, that
is the finiteness of an invariant is not a proof that there is
not a physical
singularity there). In standard black-hole
literature of curvature-formulated gravity (either General Relativity or
its modifications), one usually studies the Ricci scalar, the Kretschmann
scalar, or other invariants constructed by the Riemann tensor and its
contractions.

In teleparallel description of gravity, one has, in principle, two
approaches of finding invariants. The first is to use the solution for the
vierbein and the Weitzenb{\"{o}}ck connection in order to calculate torsion
invariants such as the torsion scalar $T$. The second is to use the
solution for the corresponding metric in order to construct the 
Levi-Civita connection, and then use it to calculate curvature invariants
such are the Ricci and Kretschmann scalars (a calculation of curvature
scalars using straightaway the Weitzenb{\"{o}}ck connection leads to zero
by construction). The comparison of both approaches is a main subject of
interest of the present work, capable of pointing out differences between
curvature and torsion gravity.

In particular, we are going to investigate whether one can formulate
everything in terms of  vierbens, Weitzenb{\"{o}}ck's connection and
torsion invariants, as one can do with the metric, the Levi-Civita
connection and curvature invariants. Perhaps one could say that the use of
curvature invariants, instead of torsion ones, is better justified by the
fact that in a realistic theory matter is coupled to the gravitational
sector through the metric and not through the vierbeins (with the
interesting
exception of fermionic matter), and particles follow geodesics defined by
the Levi-Civita connection. On the other hand, one could say that the two
approaches are equivalent only with a suitable, non-diagonal, relation
between the vierbeins and the metric. In any case, while at the classical
level the above approaches could look equivalent or alternative, for the
quantization procedure it would be crucial to determine  whether the metric
or the vierbein is the fundamental field. Therefore, the
following analysis can enlighten this subject.

The torsion invariant $T$, that is the torsion scalar, that arises from the
vierbein solution (\ref{diagonal}) with the use of Weitzenb{\"{o}}ck's
connection is  (\ref{scalartorsion1}), which in the examined case
becomes just  (\ref{TUV}). The curvature invariants that
arise from the metric solution (\ref{metric}) through the calculation of
the Levi-Civita connection are
\begin{eqnarray}\label{Ricci}
\nonumber&&R=-2\frac{G\left( r\right) ^{2}F^{\prime \prime }\left( r\right)
}{F\left(
r\right) }-2\frac{G\left( r\right) G^{\prime }\left( r\right) F^{\prime
}\left( r\right) }{F\left( r\right) }-2\left( D-2\right) \frac{G\left(
r\right) ^{2}F^{\prime }\left( r\right) }{rF\left( r\right) }\\
&& \ \ \ \ \ \ -2\left(D-2\right) \frac{G\left( r\right) G^{\prime }\left(
r\right)
}{r}-\left(
D-2\right) \left( D-3\right) \frac{G\left( r\right) ^{2}}{r^{2}}~,
\end{eqnarray}
\begin{eqnarray}\label{RR}
\nonumber &&R_{\mu \nu }R^{\mu \nu }=\frac{G\left( r\right) ^{2}}{F\left(
r\right)
^{2}r^{2}}\left[ rF^{\prime \prime }\left( r\right) G\left( r\right)
+rF^{\prime }\left( r\right) G^{\prime }\left( r\right) +\left( D-2\right)
G\left( r\right) F^{\prime }\left( r\right) \right] ^{2}\\
\nonumber
&& \ \ \ \ \ \  \ \ \ \ \ \ \ \ +\frac{G\left(
r\right) ^{2}}{F\left( r\right) ^{2}r^{2}}\left[ rF^{\prime \prime }\left(
r\right) G\left( r\right) +rF^{\prime }\left( r\right) G^{\prime }\left(
r\right) +\left( D-2\right) G^{\prime }\left( r\right) F\left( r\right)
\right] ^{2}\\
&& \ \ \ \ \ \ \ \ \   +\left( D-2\right) \frac{G\left( r\right)
^{2}}{F\left( r\right)
^{2}r^{4}}\left[ rG\left( r\right) F^{\prime }\left( r\right) +rG^{\prime
}\left( r\right) F\left( r\right) +\left( D-3\right) G\left( r\right)
F\left( r\right) \right] ^{2},
\end{eqnarray}
\begin{eqnarray}\label{RRRR}
\nonumber && R_{\mu \nu \rho \sigma }R^{\mu \nu \rho \sigma
}=4\frac{G\left( r\right) ^{2}}{F\left( r\right) ^{2}}\left[ F^{\prime
\prime }\left( r\right) G\left(
r\right) +F^{\prime }\left( r\right) G^{\prime }\left( r\right) \right]
^{2}+4\left( D-2\right) \frac{G\left( r\right) ^{4}F^{\prime }\left(
r\right) ^{2}}{r^{2}F\left( r\right) ^{2}}\\
&& \ \ \ \ \ \  \ \ \ \   \ \ \ \ \ \ \ \ \ +4\left( D-2\right)
\frac{G^{\prime}\left( r\right) ^{2}G\left( r\right)
^{2}}{r^{2}}+2\left(
D-2\right) \left( D-3\right) \frac{G\left( r\right) ^{4}}{r^{4}}~,
\end{eqnarray}
being respectively the Ricci scalar, the Ricci tensor square and the
Kretschmann scalar. Note that, using (\ref{scalartorsion1}), the Ricci
scalar is given by 
\begin{eqnarray}
\label{Ricci2}
\nonumber&&R=-T-2\frac{G\left( r\right) ^{2}F^{\prime \prime }\left(
r\right)
}{F\left(
r\right) }-2\frac{G\left( r\right) G^{\prime }\left( r\right) F^{\prime
}\left( r\right) }{F\left( r\right) } -2\left(D-2\right) \frac{G\left(
r\right) G^{\prime }\left(
r\right)
}{r}~,
\end{eqnarray}
which is just relation (\ref{RTrelation}) calculated for the vierbeins 
(\ref{diagonal}).

Observing the form of the torsion scalar $T$ in (\ref{TUV}) we
deduce that in the charged case it diverges only at $r=0$. This can be
alternatively verified examining the form  (\ref{scalartorsion1}) along
with  the expressions (\ref{G2}),(\ref{Fsol2}) (or
(\ref{G2spec}),(\ref{Fsol2spec}) for the special solution).

Observing the forms of Ricci and Kretschmann scalars in 
(\ref{Ricci}), (\ref{RRRR}) we deduce that in the charged case the possible
divergence points are at  $r=0$, at the points where
$G(r)^2\rightarrow\infty$, or at the roots of $F(r)$. From the solutions
for $G(r)$ and $F(r)$ of (\ref{G2}),(\ref{Fsol2}) we straightforwardly
obtain that $r=0$ indeed leads to divergent Ricci and Kretschmann scalars.
From the form of $G(r)$ in (\ref{G2}) along with (\ref{dfdTUV}) we observe
that $G(r)^2\rightarrow\infty$ at $1+\frac{df}{dT}=0$, that is at 
\begin{equation}\label{rs}
r_s=\left(-\frac{1+8\alpha\Lambda}{2\alpha Q^2}\right)^{\frac{1}{4-2D}}~,
\end{equation}
a relation which holds only for the negative branch, since for the
positive branch $1+\frac{df}{dT}$ has no roots (additionally since
$\alpha<0$ the above solution is real only for $\Lambda<-1/(8\alpha)$).
Indeed one can straightforwardly see that the  Ricci and Kretschmann
scalars do diverge at $r=r_s$. Finally, concerning the roots of
$F(r)$, due to (\ref{expression}), namely $F\left(r\right) =G\left(
r\right) \left( 1+\frac{df}{dT}\right)$, they are just the roots of
$G(r)$, since $F(r)$ remains finite and non-zero at $1+\frac{df}{dT}=0$
since at this point $G(r)^2\rightarrow\infty$. Taking the corresponding
limits and using (\ref{G2}),(\ref{Fsol2}), we can see that the roots of
$G(r)$ leads always to finite Ricci and Kretschmann scalars. All the
above hold also for the special solution (\ref{G2spec}),(\ref{Fsol2spec}).
In summary, in the charged case the Ricci and Kretschmann scalars diverge
at $r=0$ and at $r=r_s$ given by (\ref{rs}). Lastly, we mention that in the
uncharged case the Ricci scalar vanishes, however the Kretschmann scalar
behaves as in the charged case.

From the above analysis we are led to the very interesting result that in
some cases the singularities obtained by the torsion scalar analysis are
less than those obtained by the curvature scalar analysis. In particular,
this happens for the negative branch of the solutions, for $Q\neq0$ and
for $\Lambda<-1/(8\alpha)$, in which case the curvature invariants possess
an additional physical singularity at $r=r_s$ given  by (\ref{rs}). We
stress that when $f(T)=0$, that is in the case of usual teleparallel
gravity, the negative branch disappears as we mentioned above, thus the
singularity analyses of the two approaches coincide. Additionally, in the
uncharged case, that is when $Q=0$, the extra singularity at $r_s$
disappears too, and the singularity analyses of the two approaches coincide
too. In conclusion, we deduce that the above difference in the physical
singularities of the torsion and curvature analysis, is a result of both
the non-linear $f(T)$ structure and of the non-zero electric charge, which
reveals the novel features that are brought in in the theory in this case.

Let us discuss on the horizons of the above solutions. Although
we showed that the roots of $G(r)$ at $r>0$ (if they exist) do not
correspond to physical
singularities, obviously they correspond to horizons, since they appear in
the denominator in the metric (\ref{metric}) and in the vierbeins
(\ref{diagonal}). In order to show that the roots of $G(r)$, say at
$r=r_H$, are just coordinate singularities, we consider the
Painlev\'{e}-Gullstrand
coordinates \cite{Painleve,
Gullstrand, Martel:2000rn, Liu:2005hj} through the transformation 
$dt=d\tau +g\left( r\right) dr$,
with $g\left( r\right) $ a function of the radial coordinate. Therefore,
the metric (\ref{metric}) becomes
\begin{equation}
ds^{2}=F\left( r\right) ^{2}d\tau ^{2}+2g\left( r\right) F\left( r\right)
^{2}drd\tau -\left[ \frac{1}{G\left( r\right) ^{2}}-F\left( r\right)
^{2}g\left( r\right) ^{2}\right]
dr^{2}-r^{2}\sum_{i=1}^{i=D-2}dx_{i}^{2}~\,.
 \label{metricPG}
\end{equation}
Choosing $g\left( r\right) ^{2}=\frac{1}{F\left( r\right) ^{2}}\left[ 
\frac{1}{G\left( r\right) ^{2}}-1\right] $ and defining 
\begin{equation}
h(r)^{2}=F\left(r\right) ^{2}/G\left( r\right) ^{2}=\left[2+\sqrt{1-24\alpha
\Lambda -6\alpha
Q^{2}r^{4-2D}}\right]/3
\end{equation}
we can bring it in a flat Euclidean form 
\begin{equation}
ds^{2}=F\left( r\right) ^{2}d\tau ^{2}+2h\left( r\right) \sqrt{1-G\left(
r\right) ^{2}}drd\tau -dr^{2}-r^{2}\sum_{i=1}^{i=D-2}dx_{i}^{2}~,
\label{metricregular}
\end{equation}
which is regular at $r=r_{H}$. Therefore, $r=r_{H}$, if they exist,  are
just coordinate singularities, that is horizons.

From the above analysis it is implied that the black-hole solutions of the
charged $f(T)$ gravity may possess a horizon at $r_H$ that shields the
physical singularities. However, firstly it is not guaranteed that $r_H$
exists, since there could be parameter choices for which $G(r)$ has no
roots, that is the physical singularity at $r=0$ becomes naked.
Secondly, even if $r_H$ exists 
it is not guaranteed that it will shield the second physical singularity of
the charged negative branch at $r=r_s$ given by (\ref{rs}), since this will
depend on the specific parameter choice. In particular, we can see that if
$F(r_s)^2<0$ then $r_H$ exists and shields the singularity at $r_s$, that
is $r_H>r_s$, otherwise $r_s$ is a naked singularity. This is not the case
for $f(T)\rightarrow0$ or  $Q\rightarrow0$, in which, as we mentioned,
$r_s$ disappears. Therefore, we conclude that the cosmic censorship
theorem, namely that there are always horizons that shield the physical
singularities, does not always hold for $f(T)$ gravity, a result that was
already found in the 3D case too \cite{Gonzalez:2011dr}.

Before proceeding to the numerical elaboration of the obtained solutions,
we make the following comment. In curvature gravity there can be cases
where the Ricci scalar is finite at one point although there is a physical
singularity there, which is revealed through the use of the Kretschmann
scalar, and that is why people usually examine both scalars
simultaneously. Thus, one could ask whether one should use additional
torsion scalars too, defined as various contractions of the torsion
tensor. In particular, according to (\ref{telelag}), the torsion scalar 
$T$ contains three separate scalars, corresponding to different
contractions of the torsion tensor, namely $I_1=T^{\mu \nu \rho }T_{\mu \nu
\rho }$, $I_2=
T^{\rho \mu \nu}
T_{\nu \mu \rho }$, $I_3=
T_{\rho \mu }^{ \ \ \ 
\rho }
T_{ \ \ \ \nu }^{\nu \mu }$, and thus one could additionally examine their
behavior in order to reveal the singularities. However, it is well
known that these separate combinations are not invariant under local
Lorentz transformations, and that was the reason that the teleparallel
Lagrangian (torsion scalar) $T$ was defined as their specific combination
which becomes Lorentz invariant \cite{Hayashi79}. Therefore, one cannot use
other torsion scalars apart from $T$ in order to investigate the
singularities (the explicit calculation of $I_i$'s for the obtained
vierbein
solutions shows that they acquire  different values in different
coordinates, and thus they are not invariants, however their specific
combination in $T$ does acquire the same value independently of the
coordinate basis and thus it is a well-defined invariant).

In order to provide a more transparent picture of the above singularity and
horizon behavior, we proceed to the numerical elaboration of specific
examples. Since the asymptotic behavior of $G(r)$ is given by 
\begin{equation}
G( r )^2=-\Lambda_{eff}\,r^2+ \ldots~,
\end{equation}
where $\ldots$ correspond to sub-leading terms and 
\begin{equation}
\label{Leff}
\Lambda_{eff}=\frac{9}{54\alpha(D-1)(D-2)(2\pm
\sqrt{1-24\alpha\Lambda})}\left[1+72\alpha\Lambda\mp
(1-24\alpha\Lambda)^{3/2}\right]~,
\end{equation}
we can distinguish three subclasses, namely the asymptotically AdS one
($\Lambda_{eff}<0$), the asymptotically dS one ($\Lambda_{eff}>0$) and the
limiting $\Lambda_{eff}=0$ one. Note that the second and third solution
subclasses can never be obtained by the negative solution branch. Without
loss of generality we consider $D=4$, $\alpha=-1$ and $Const =-1$, while we
suitably choose $\Lambda$ in order to lie in the above three subclasses,
which we investigate separately.\\

\begin{itemize}

\item {\bf{Case $\Lambda_{eff}<0$}}

Let us  consider $\Lambda=-1/25$,  which satisfies the
condition $\Lambda_{eff}<0$. In Fig.~\ref{CaseAdS} from up to down we
depict $G(r)^2$, the Ricci scalar, the Kretschmann scalar and the torsion
scalar $T(r)$ as a function of $r$, for the positive and negative branches
of black-hole solutions. The left graphs correspond to charged solutions
($Q=1$), while the right graphs correspond to uncharged solutions ($Q=0$).
\begin{figure}[!]
\begin{center}
\includegraphics[width=2.8in,angle=0,clip=true]{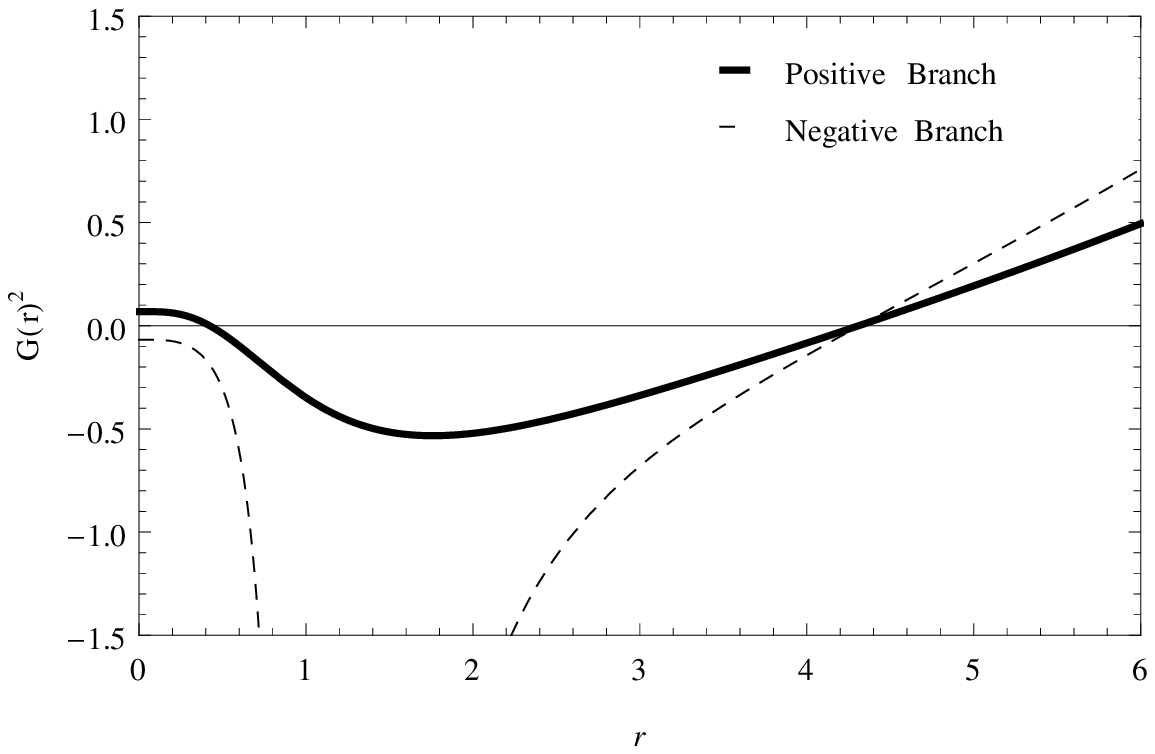}
\includegraphics[width=2.8in,angle=0,clip=true]{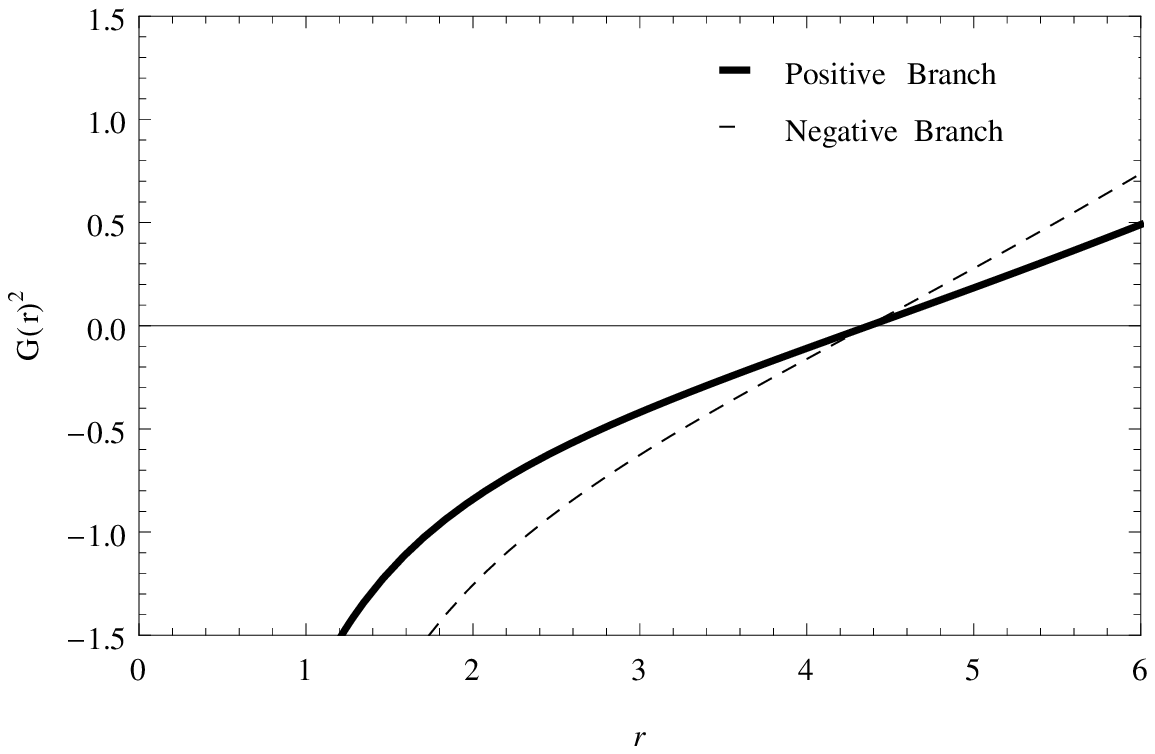}
\includegraphics[width=2.8in,angle=0,clip=true]{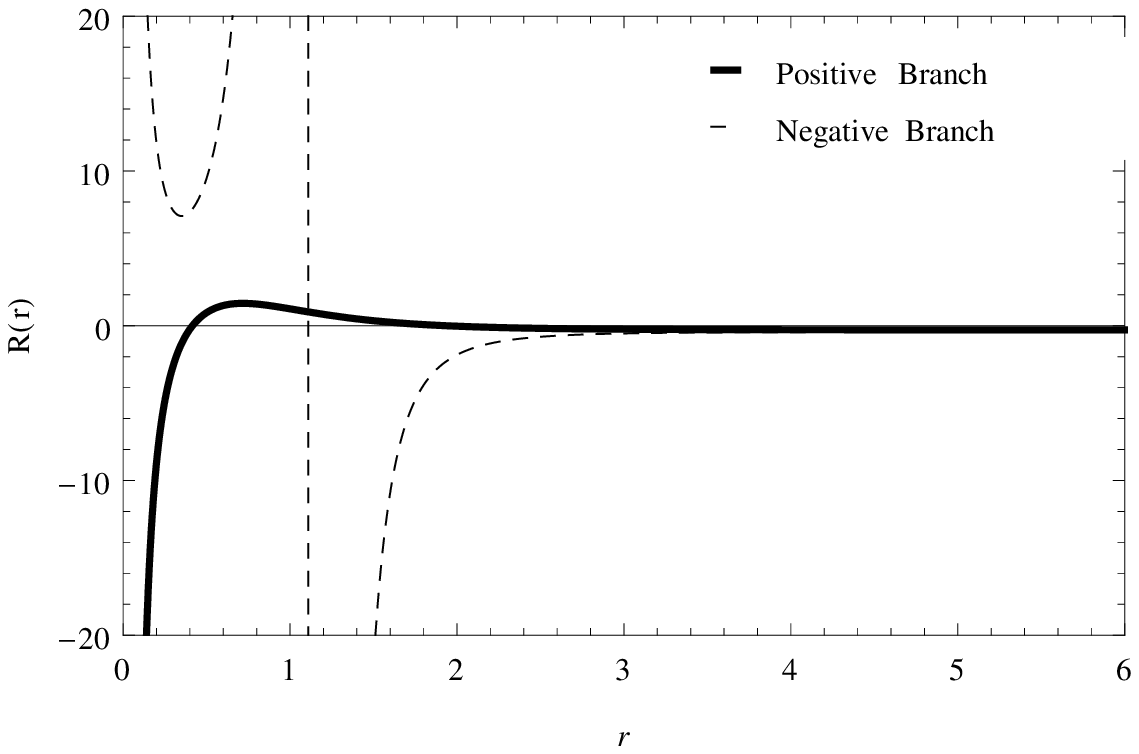}
\includegraphics[width=2.8in,angle=0,clip=true]{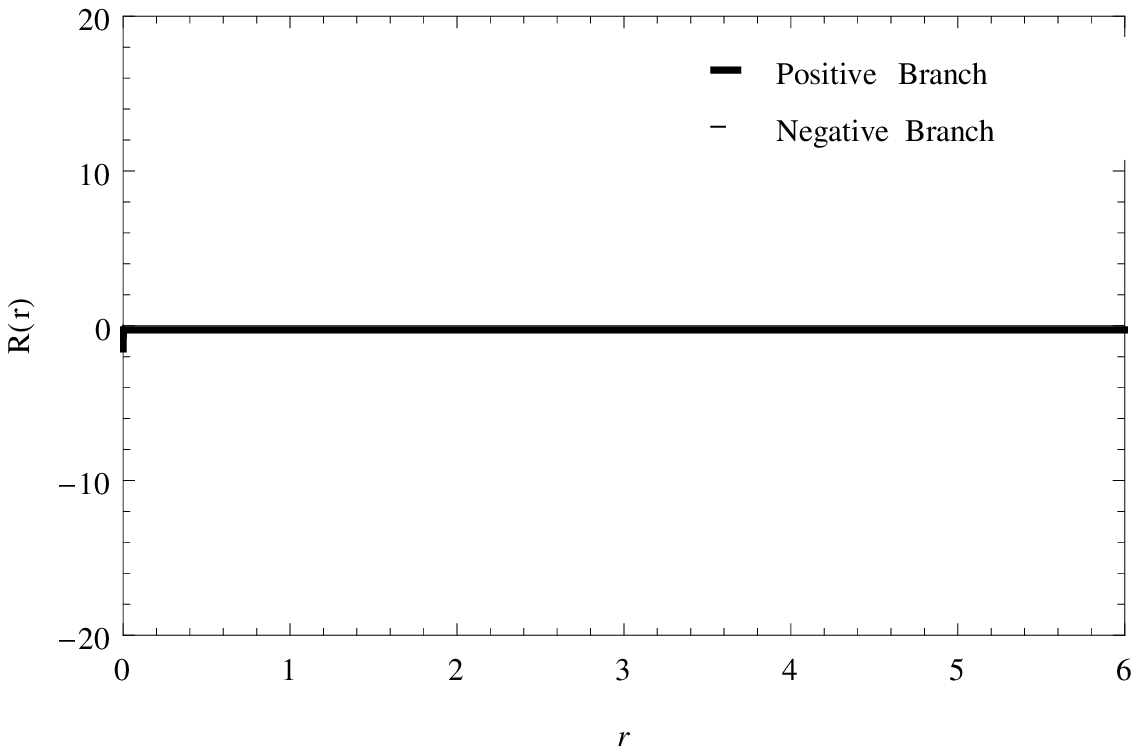}
\includegraphics[width=2.8in,angle=0,clip=true]{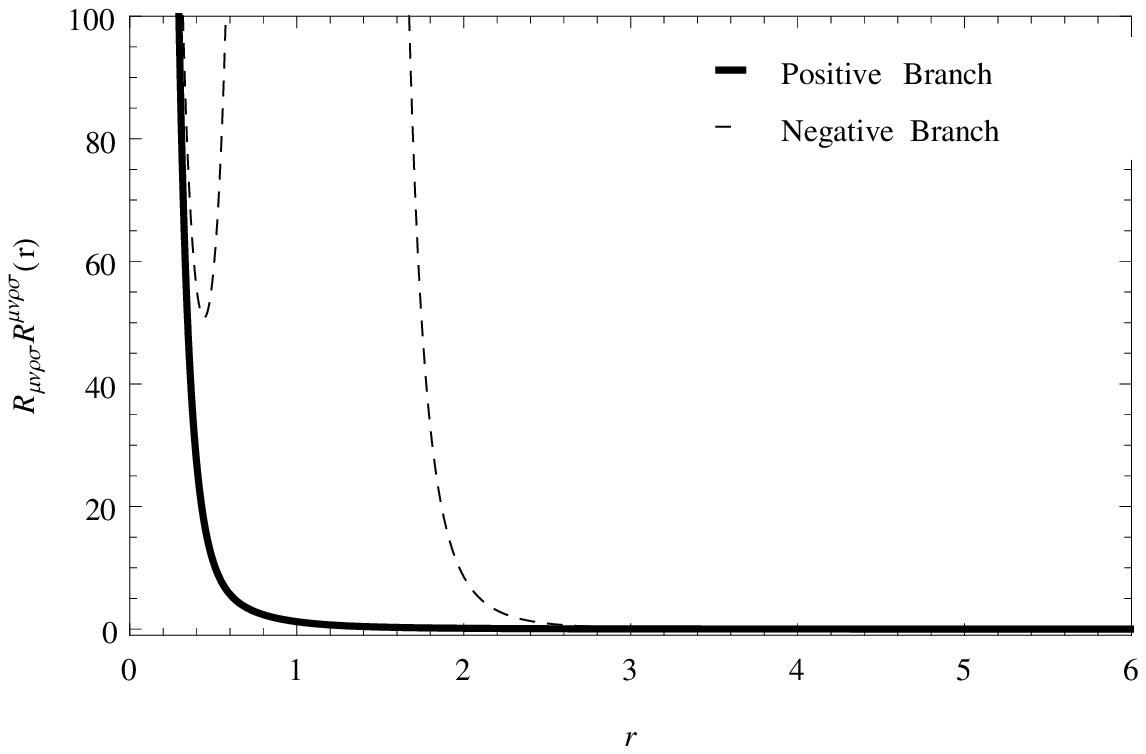}
\includegraphics[width=2.8in,angle=0,clip=true]{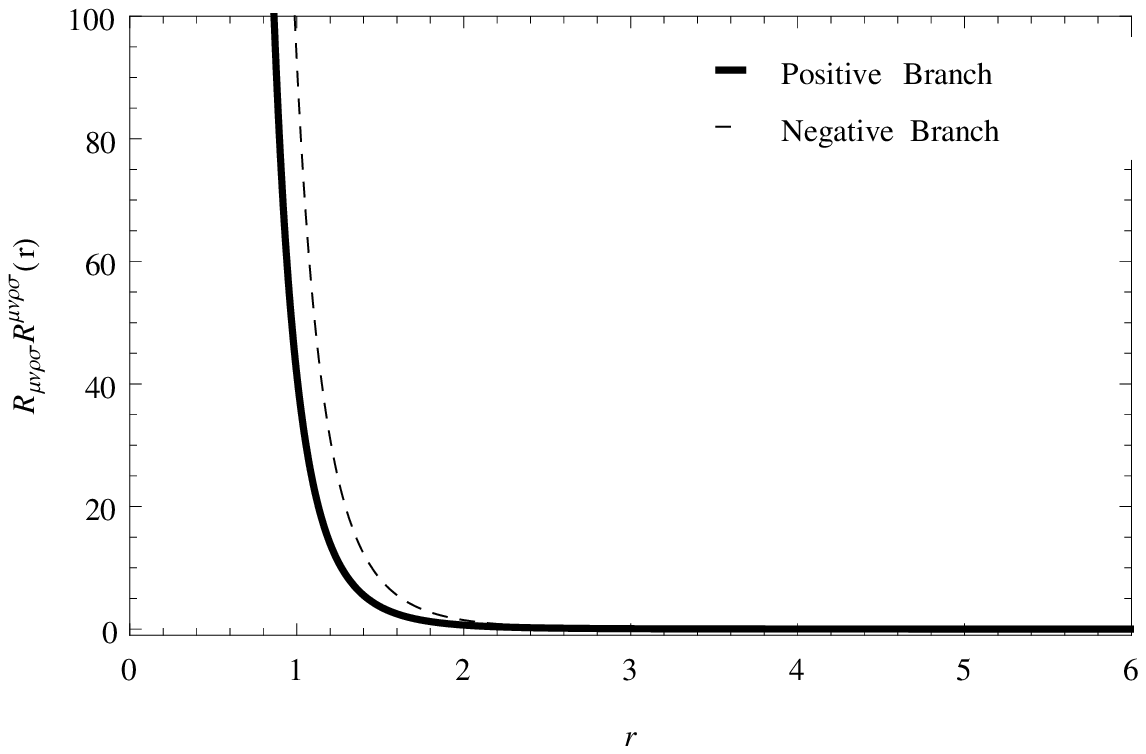}
\includegraphics[width=2.8in,angle=0,clip=true]{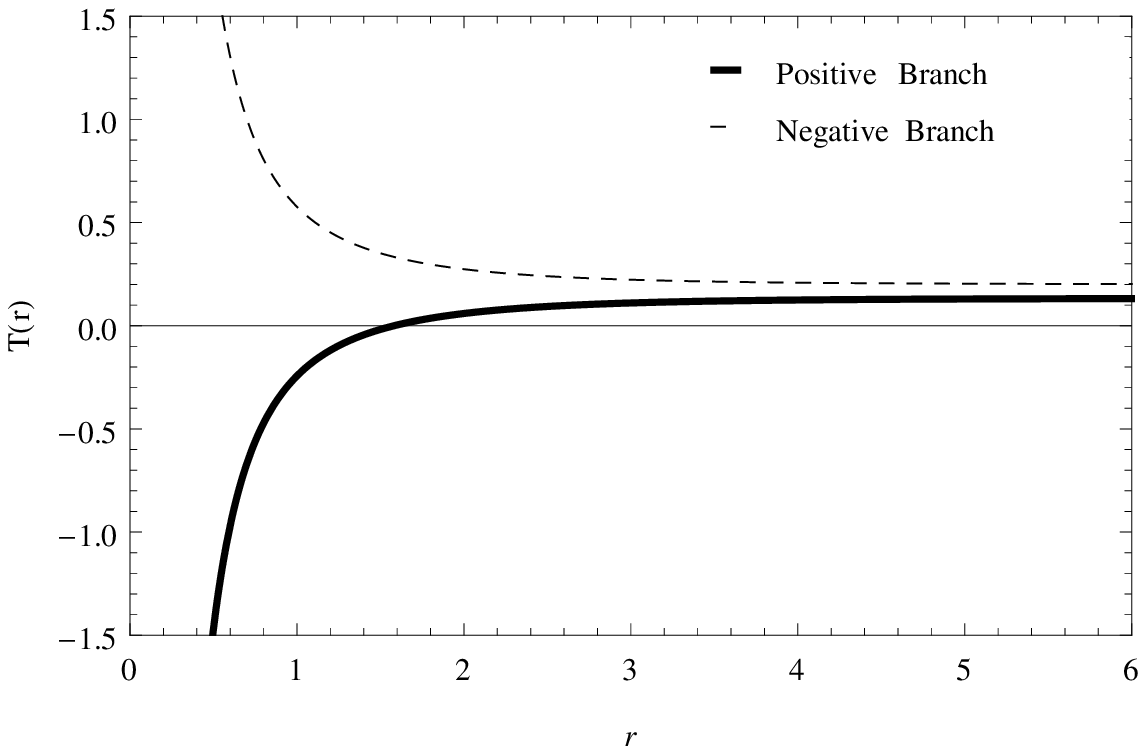}
\includegraphics[width=2.8in,angle=0,clip=true]{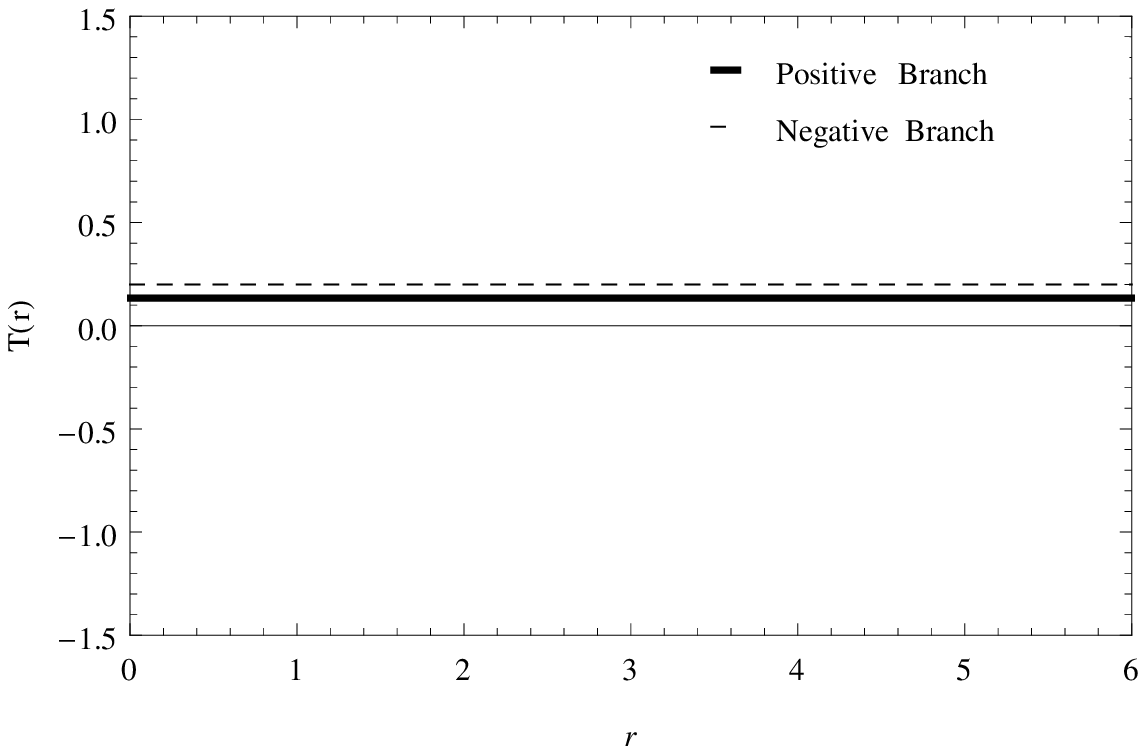}
\caption{{\it{The solutions for $G( r )^2$ of (\ref{G}), for the Ricci
scalar $R(r)$ of (\ref{Ricci}), for the Kretschmann scalar $R_{\mu \nu \rho
\sigma }R^{\mu \nu \rho \sigma }(r)$  of (\ref{RRRR}) and for the torsion
scalar $T(r)$ of (\ref{scalartorsion1}), as a function of $r$, for the
positive
(thick solid curve) and negative (thin dashed curve) branch of the AdS
solution subclass, for  $D=4$,
$\alpha=-1$, $Const =-1$ and $\Lambda=-1/25$. Left
graphs correspond
to charged solutions with $Q=1$, while right graphs correspond to
uncharged solutions with $Q=0$. The thin $0$-line is depicted for
convenience.}}}
\label{CaseAdS}
\end{center}
\end{figure}

The positive branch in the charged case exhibits only one physical
singularity at $r=0$, in which the torsion scalar and the Ricci and
Kretschmann scalars diverge, however it is shielded by two horizons at
$r_H=r_-$ and $r_H=r_+$, since in this case $G(r)^2$ has two roots. In
order to examine whether $r_{+}$ is a Killing horizon we see that the
timelike Killing vector of the metric is $\epsilon ^{\mu}\partial _{\mu
}=\partial _{t}$ \footnote{Note that none of the metric coefficients
depends on time and thus the manifold has a timelike Killing vector
$\partial _{t}$, and similarly since none of the metric coefficients
depends on $x_i $ there exist $(D-2)$ spacelike Killing vector fields
$\partial _{x_i }$.},
with norm $\epsilon _{\mu }\epsilon ^{\mu }=g_{tt}=F\left( r\right) ^{2}$
which vanishes at $r=r_{+}$. Inside the horizon the Killing vector field
is spacelike, while outside it is timelike, and thus it corresponds to a
null hypersurface. Finally, for the uncharged solutions we can see that
the Ricci scalar and the torsion scalar are constants, while the
Kretschmann scalar diverge at $r=0$ (similarly to usual
General Relativity).

For the negative branch in the charged case, the torsion scalar possesses
only one divergence, namely at $r=0$, however the Ricci and Kretschmann
scalars possess two divergence points, namely at $r=0$ and $r=r_s$ as
described above. However, in this specific numerical example, both these
physical singularities are shielded by the horizon at $r_H>r_s$, in which
all invariants remain regular. In particular, $r_H$ is a Killing
horizon, corresponding to an event horizon since the Killing vector field
is timelike outside the horizon and spacelike inside. Finally, 
for the uncharged solutions the Ricci  and torsion scalars are
constants, but the Kretschmann scalar diverges at $r=0$. However, this
physical singularity is shielded by the horizon at $r=r_H$ in which
$G(r)^2$ becomes zero.

In summary, we indeed verify that in the charged case and for the negative
branch the curvature invariants contain an extra divergence at $r=r_s$,
that does not appear in the torsion invariant, revealing the novel
features of charged $f(T)$ gravity.

\item {\bf{Case $\Lambda_{eff}>0$}}
  
We consider $\Lambda=1/25$, which satisfies the condition $\Lambda_{eff}>0$
for the positive branch (as we mentioned below (\ref{Leff}) the
negative branch cannot lead to $\Lambda_{eff}>0$). In Fig.~\ref{CasedS}
from up to down we depict $G(r)^2$, the Ricci scalar, the Kretschmann
scalar and the torsion scalar $T(r)$ as a function of $r$, for the positive
branch of black-hole solutions. The left graphs correspond to charged
solutions ($Q=1$), while the right graphs correspond to uncharged solutions
($Q=0$). 
\begin{figure}[!]
\begin{center}
\includegraphics[width=2.7in,angle=0,clip=true]{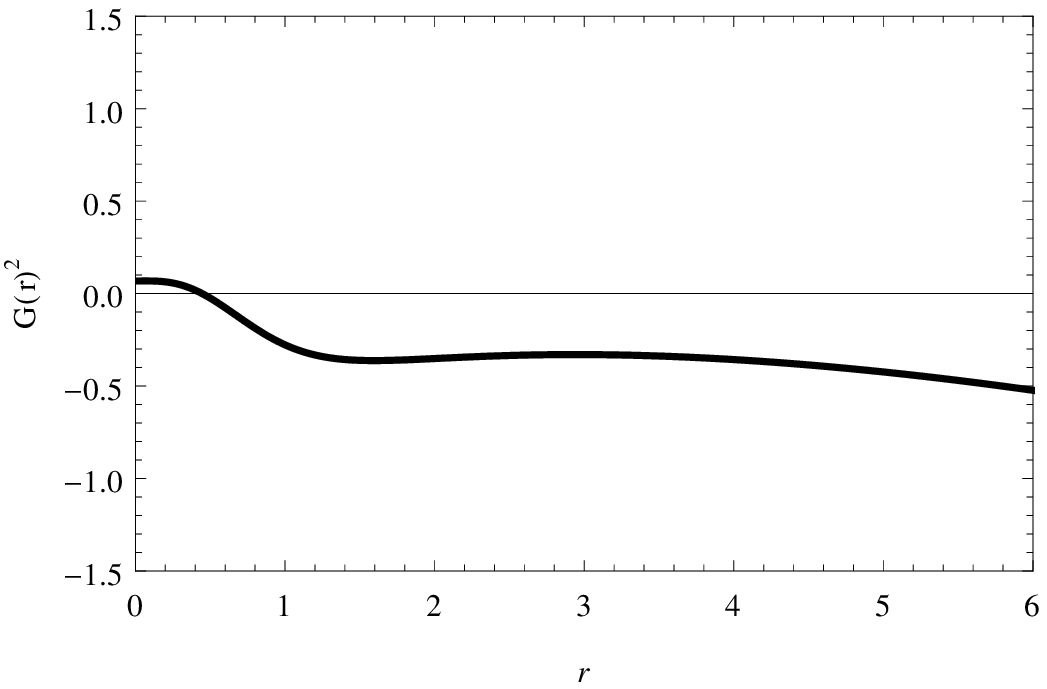}
\includegraphics[width=2.7in,angle=0,clip=true]{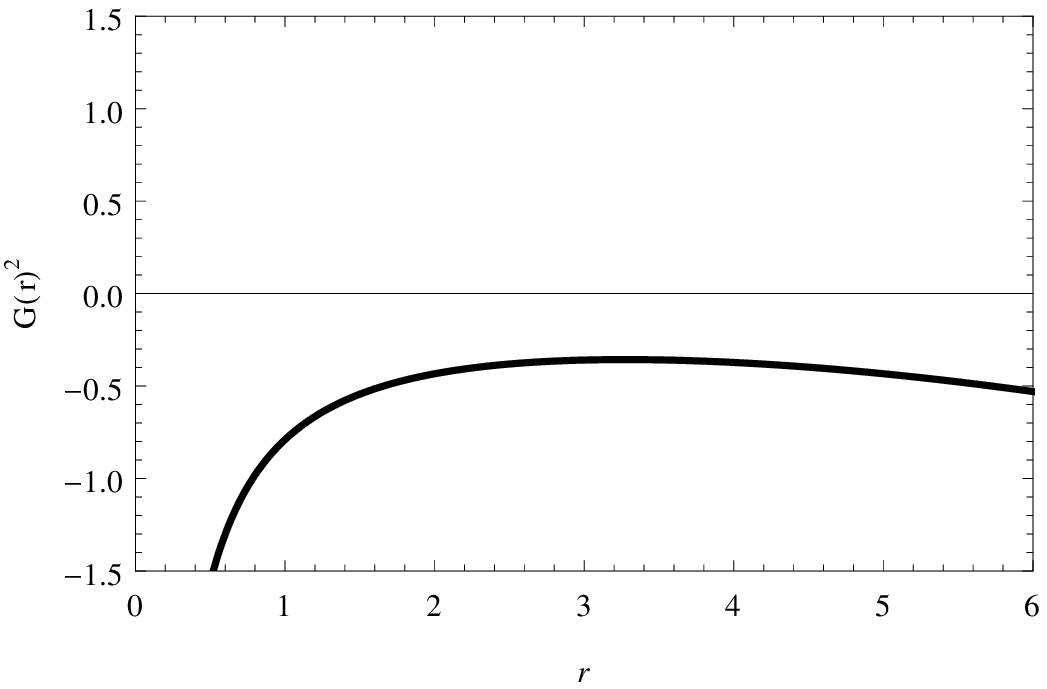}
\includegraphics[width=2.7in,angle=0,clip=true]{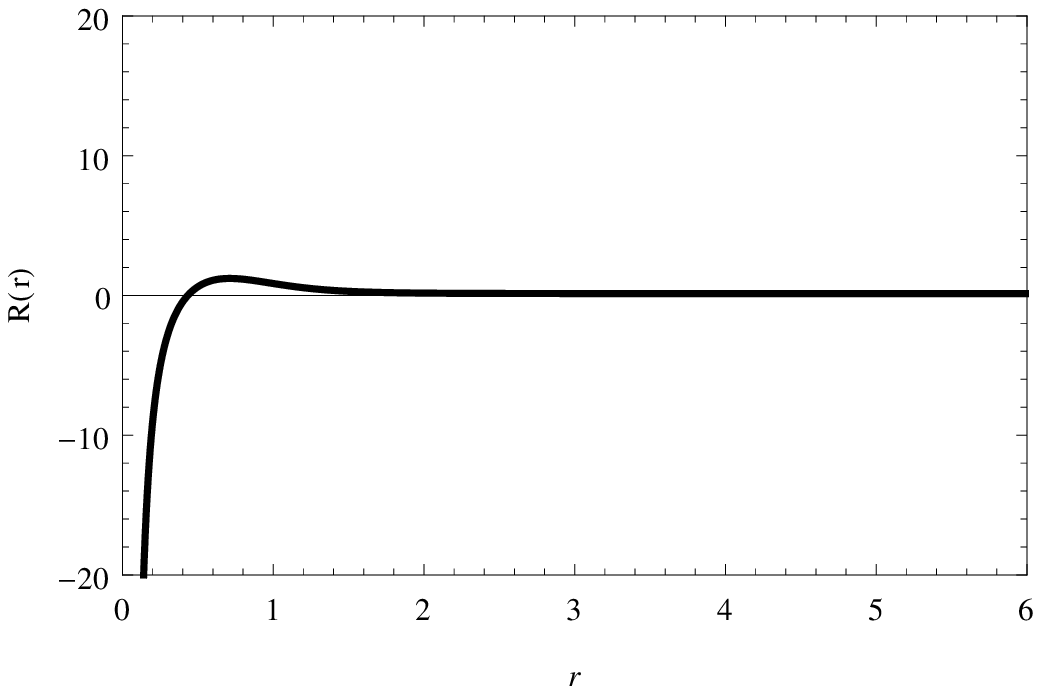}
\includegraphics[width=2.7in,angle=0,clip=true]{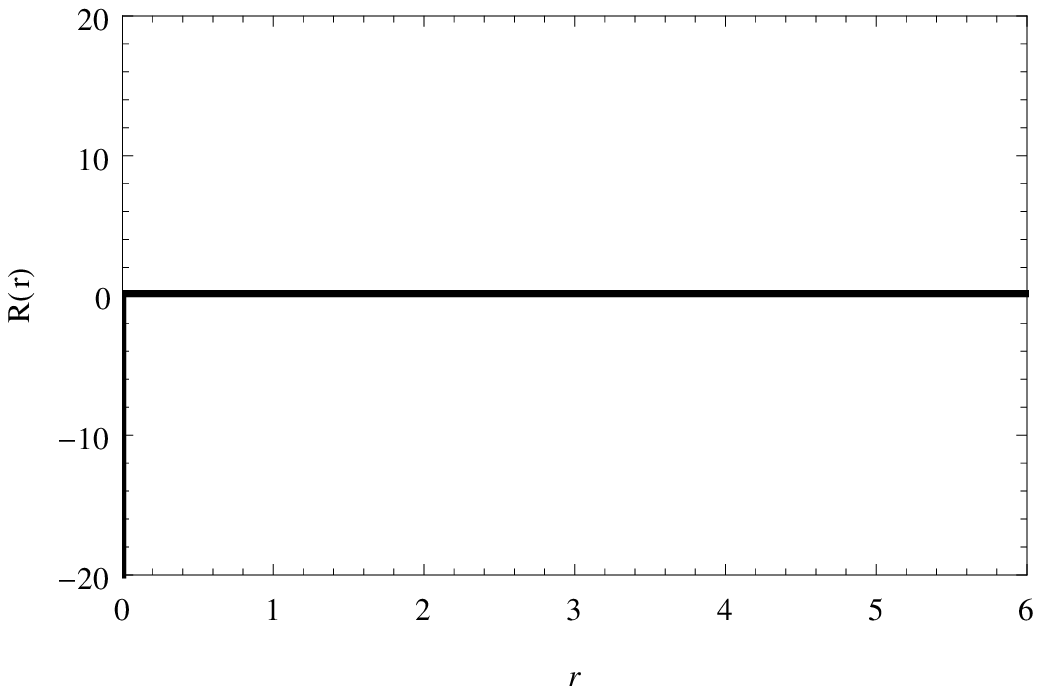}
\includegraphics[width=2.7in,angle=0,clip=true]{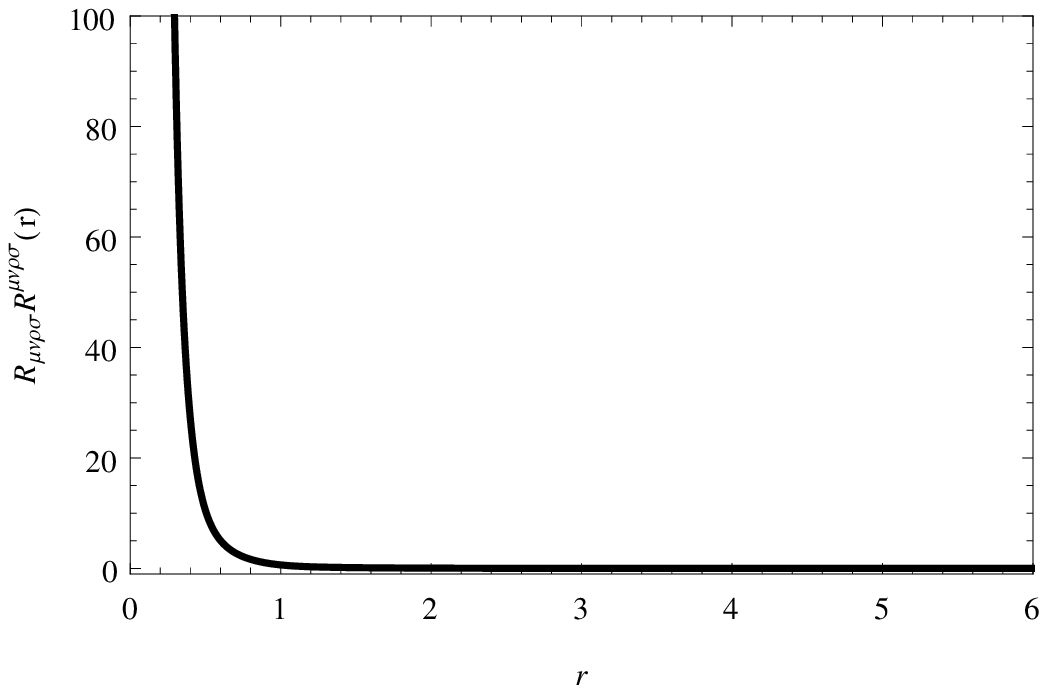}
\includegraphics[width=2.7in,angle=0,clip=true]{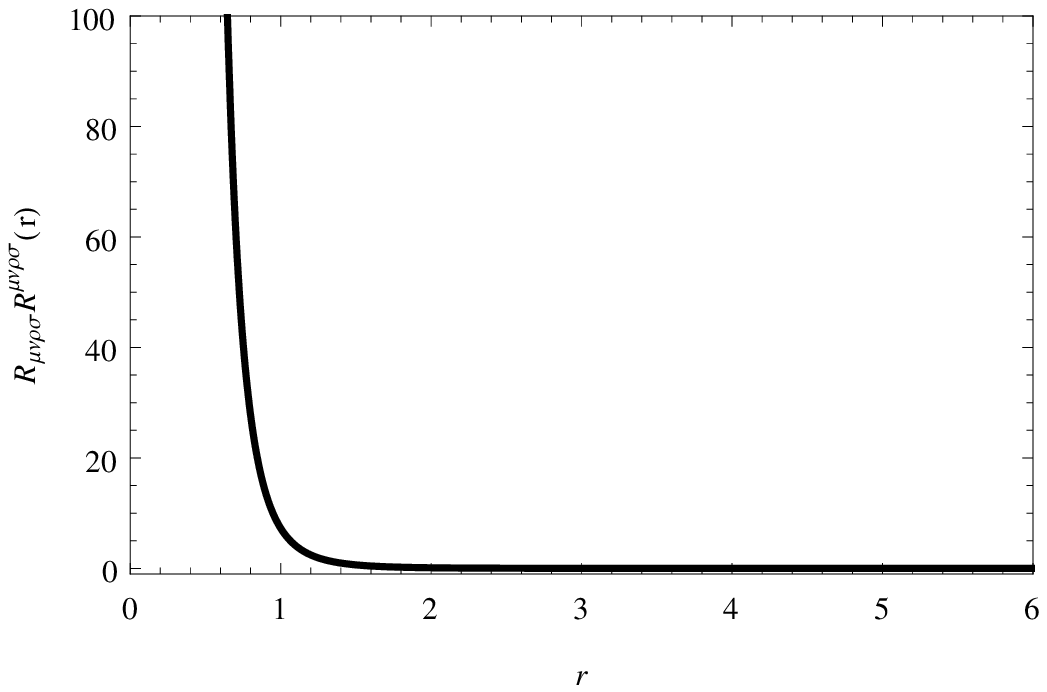}
\includegraphics[width=2.7in,angle=0,clip=true]{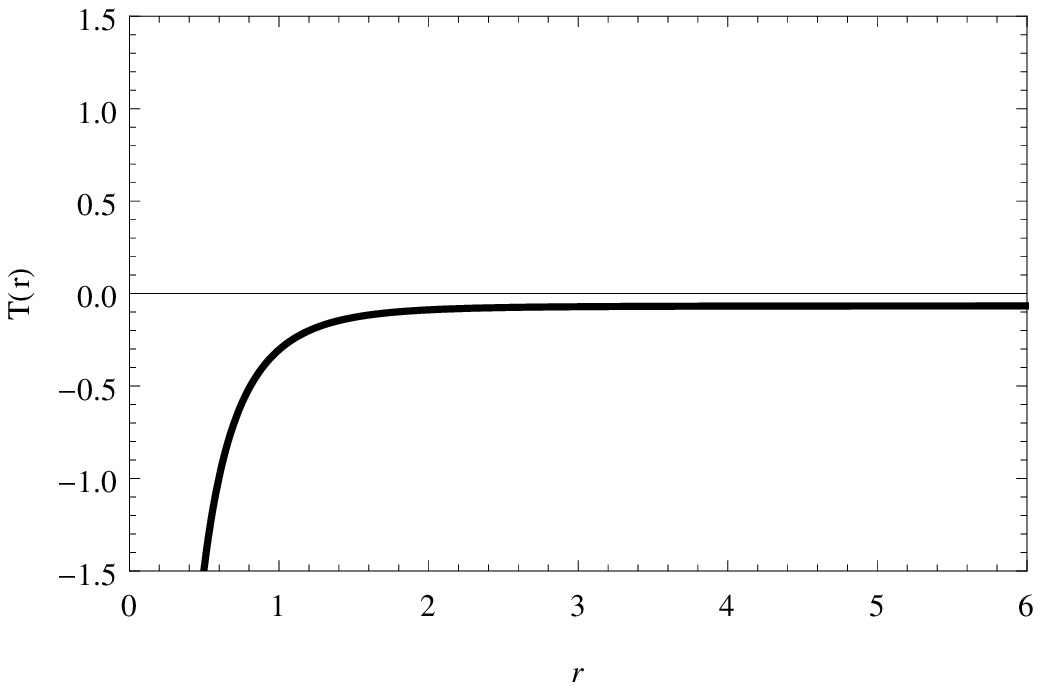}
\includegraphics[width=2.7in,angle=0,clip=true]{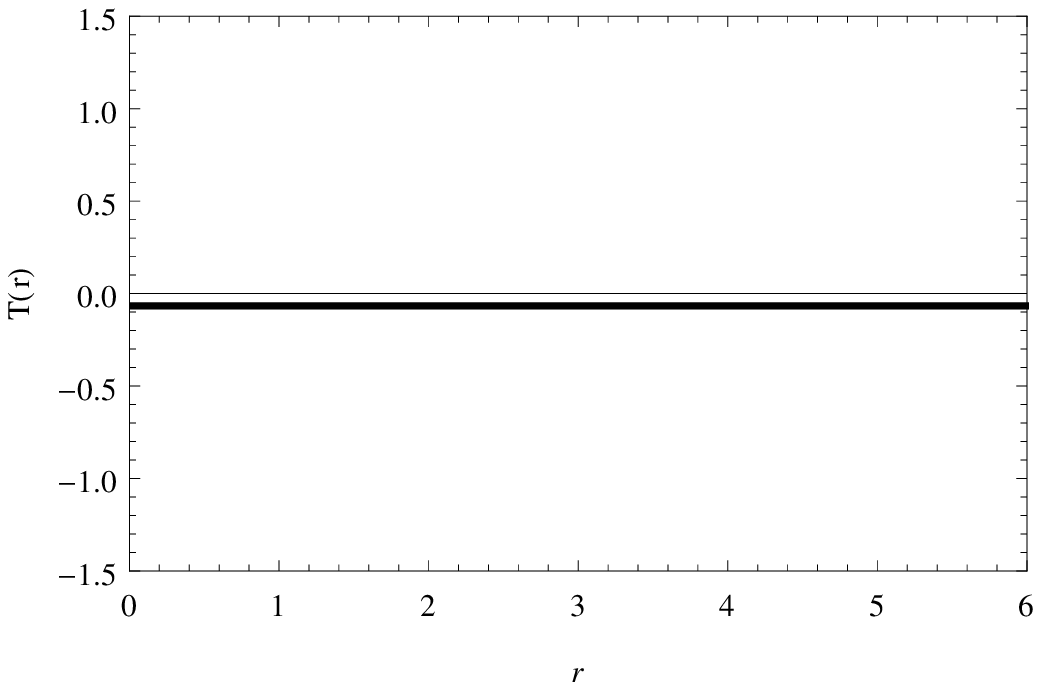}
\caption{{\it{The solutions for $G( r )^2$ of (\ref{G}), for the Ricci
scalar $R(r)$ of (\ref{Ricci}), for the Kretschmann scalar $R_{\mu \nu \rho
\sigma }R^{\mu \nu \rho \sigma }(r)$  of (\ref{RRRR}) and for the torsion
scalar $T(r)$ of (\ref{scalartorsion1}), as a function of $r$, for the
positive
(thick solid curve) branch of the dS
solution subclass, for  $D=4$,
$\alpha=-1$, $Const =-1$ and $\Lambda=1/25$. Left graphs correspond
to charged solutions with $Q=1$, while right graphs correspond to
uncharged solutions with $Q=0$. The thin $0$-line is depicted for
convenience.}}}
\label{CasedS}
\end{center}
\end{figure}

The charged case possesses a physical singularity at $r=0$, where the
torsion scalar and the Ricci and Kretschmann scalars diverge, however it is
shielded a horizon at $r=r_H$, where $G(r)^2$ becomes zero. 
In order to examine whether $r_H$ is a Killing horizon, and similarly to
the previous case, we observe that the timelike Killing vector of the
metric is $\epsilon ^{\mu}\partial _{\mu }=\partial _{t}$,
with norm $\epsilon _{\mu }\epsilon ^{\mu }=g_{tt}=F\left( r\right) ^{2}$
which vanishes at $r=r_{H}$. Since outside the horizon the Killing
vector field is timelike, and inside it is spacelike, it is implied that it
corresponds to a null hypersurface, that is a cosmological Killing horizon.

In the uncharged case we observe that the Ricci scalar vanishes while the
torsion scalar is constant, however the Kretschmann scalar diverges at
$r=0$. However, note that in this case $G(r)^2$ has no roots, that is
there is not a horizon to shield the physical singularity at $r=0$, which
is therefore a naked one.

\item {\bf{Case $\Lambda_{eff}=0$}}

\begin{figure}[!]
\begin{center}
\includegraphics[width=2.7in,angle=0,clip=true]{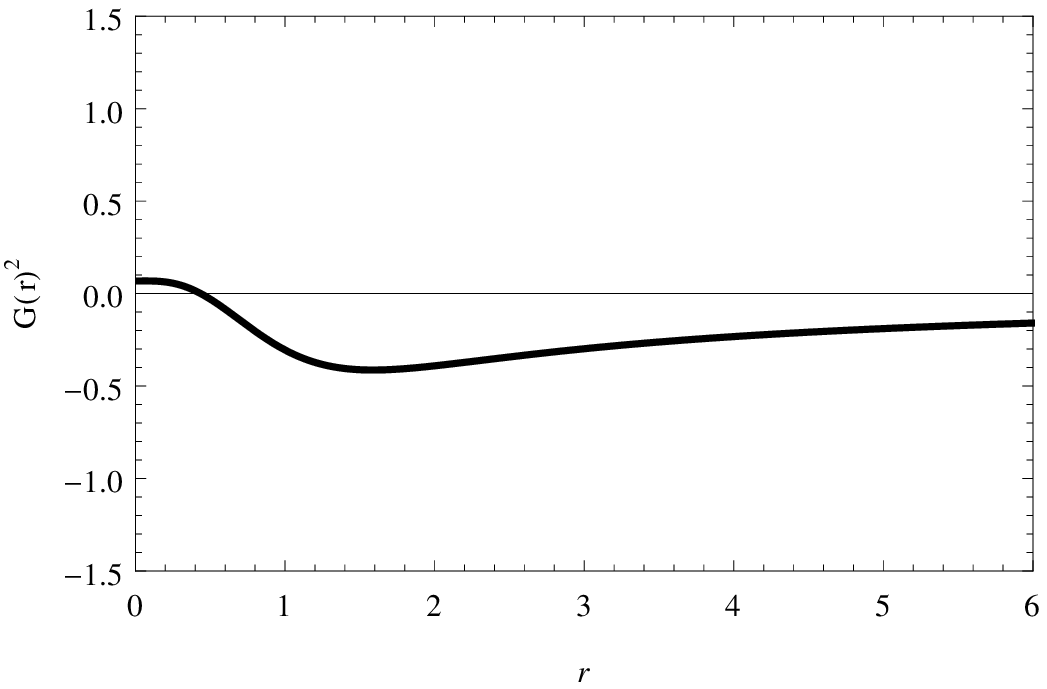}
\includegraphics[width=2.7in,angle=0,clip=true]{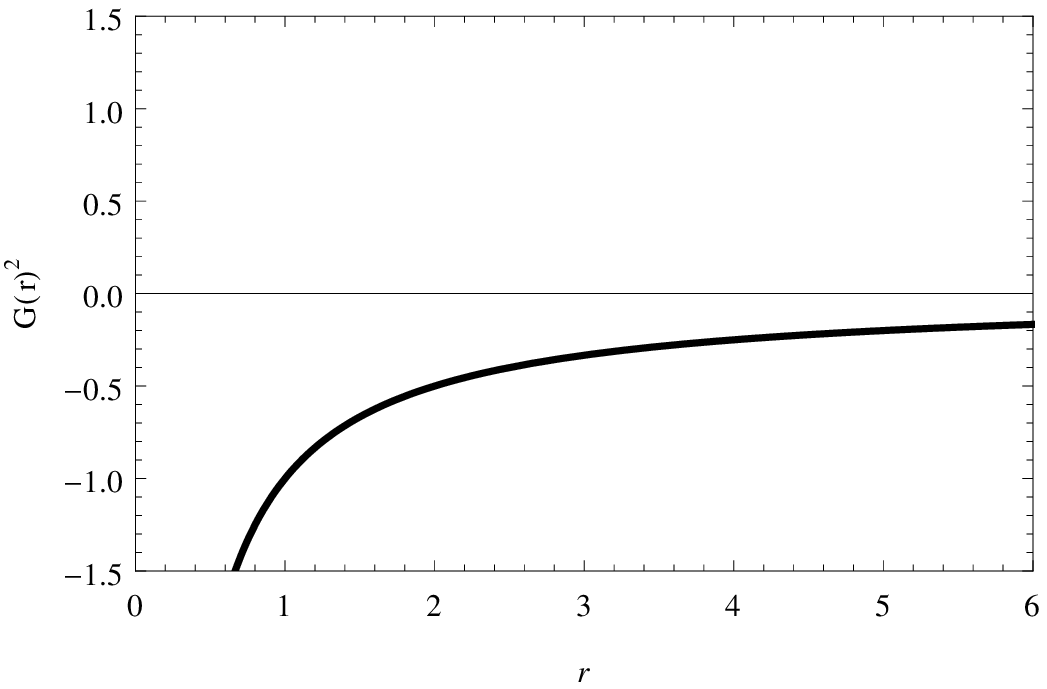}
\includegraphics[width=2.7in,angle=0,clip=true]{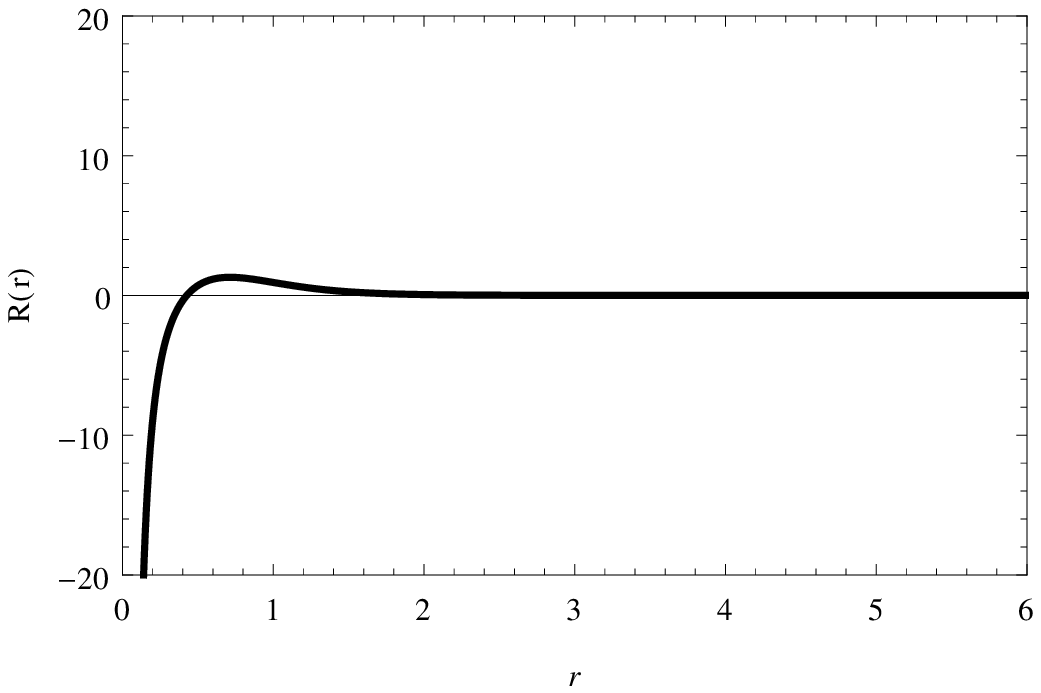}
\includegraphics[width=2.7in,angle=0,clip=true]{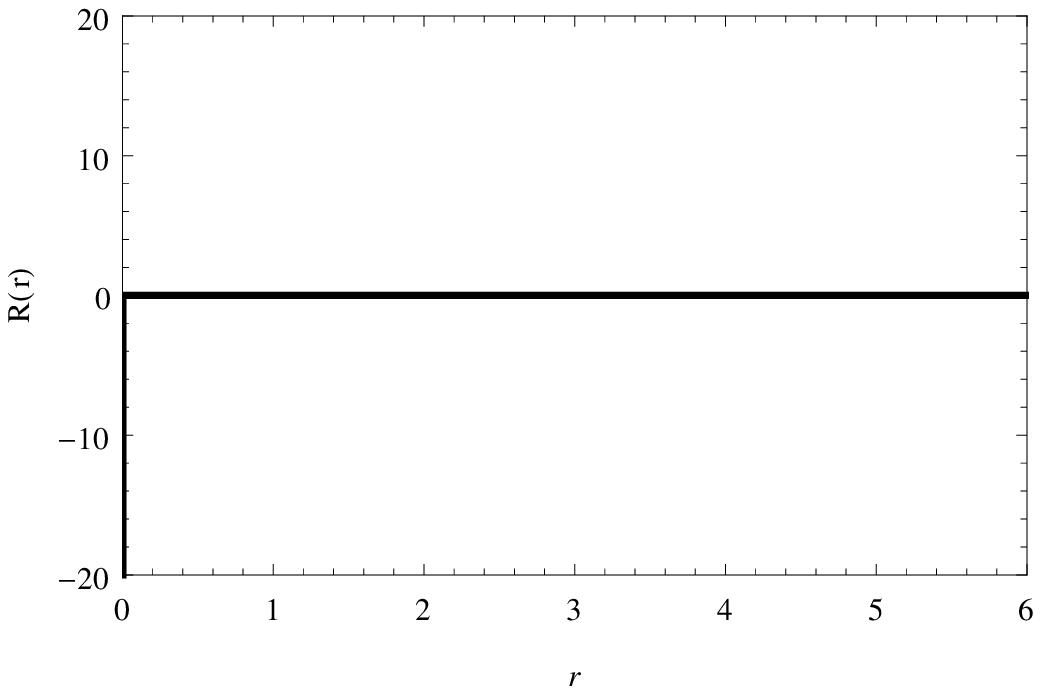}
\includegraphics[width=2.7in,angle=0,clip=true]{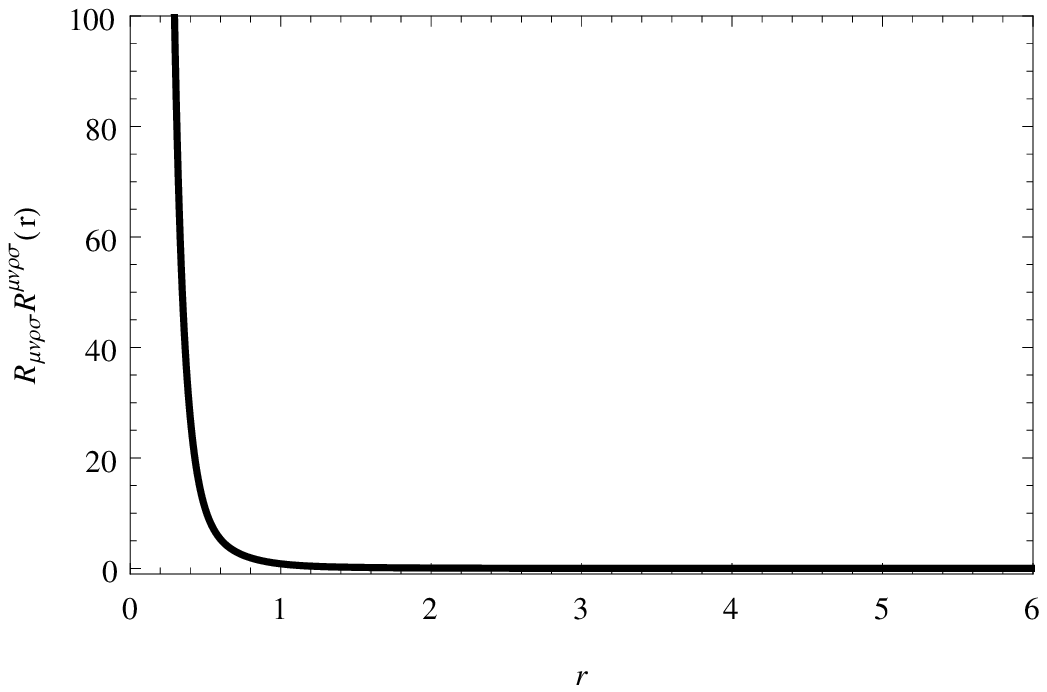}
\includegraphics[width=2.7in,angle=0,clip=true]{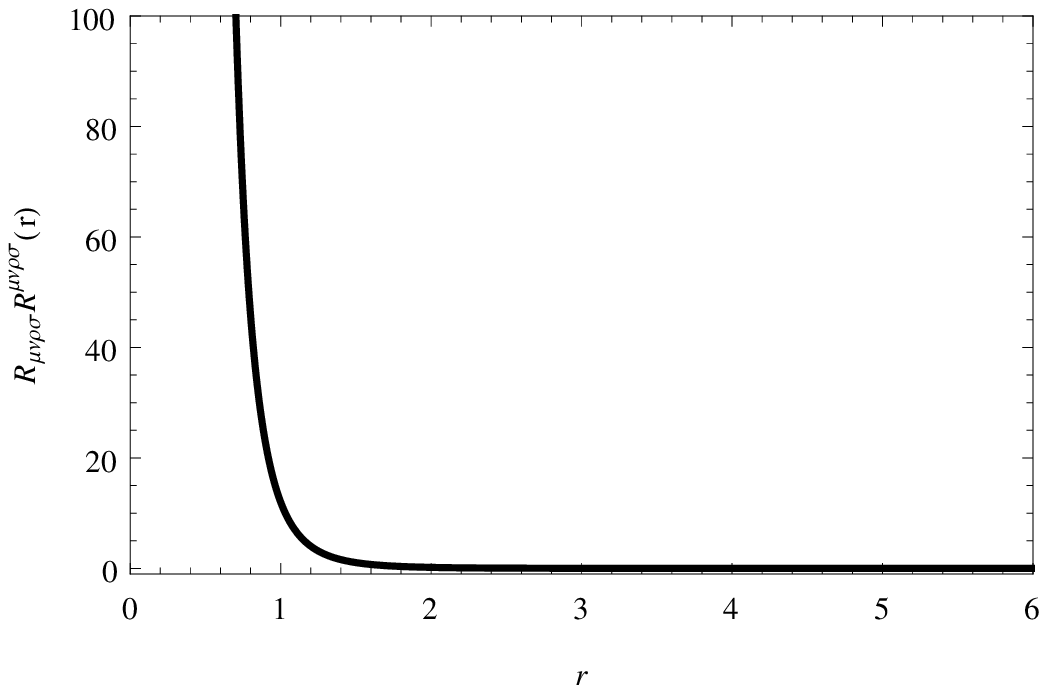}
\includegraphics[width=2.7in,angle=0,clip=true]{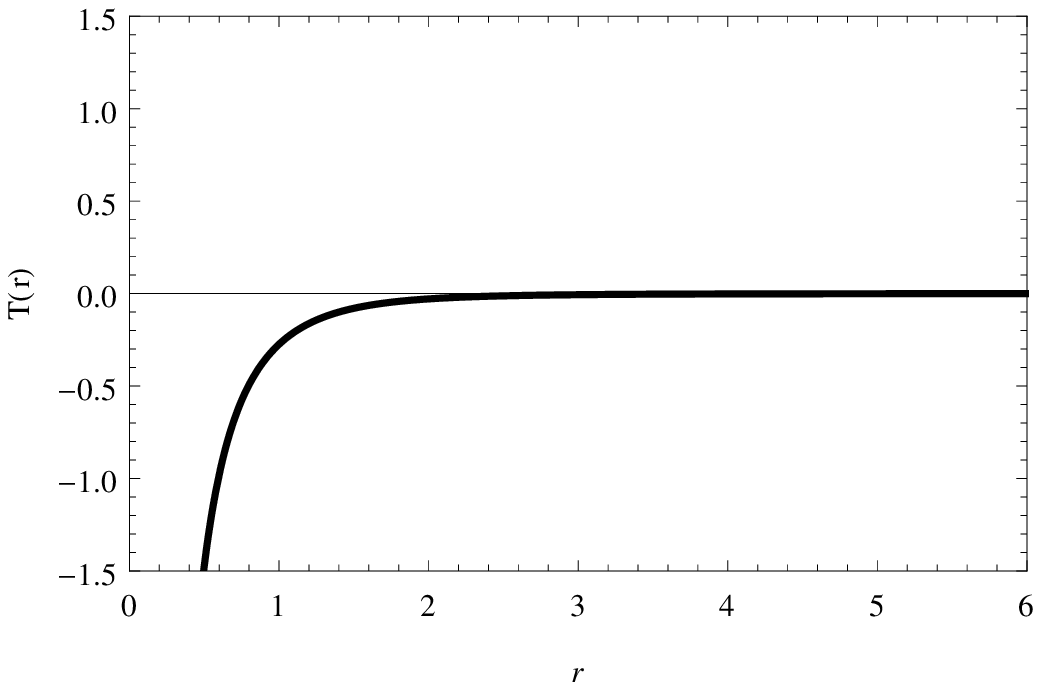}
\includegraphics[width=2.7in,angle=0,clip=true]{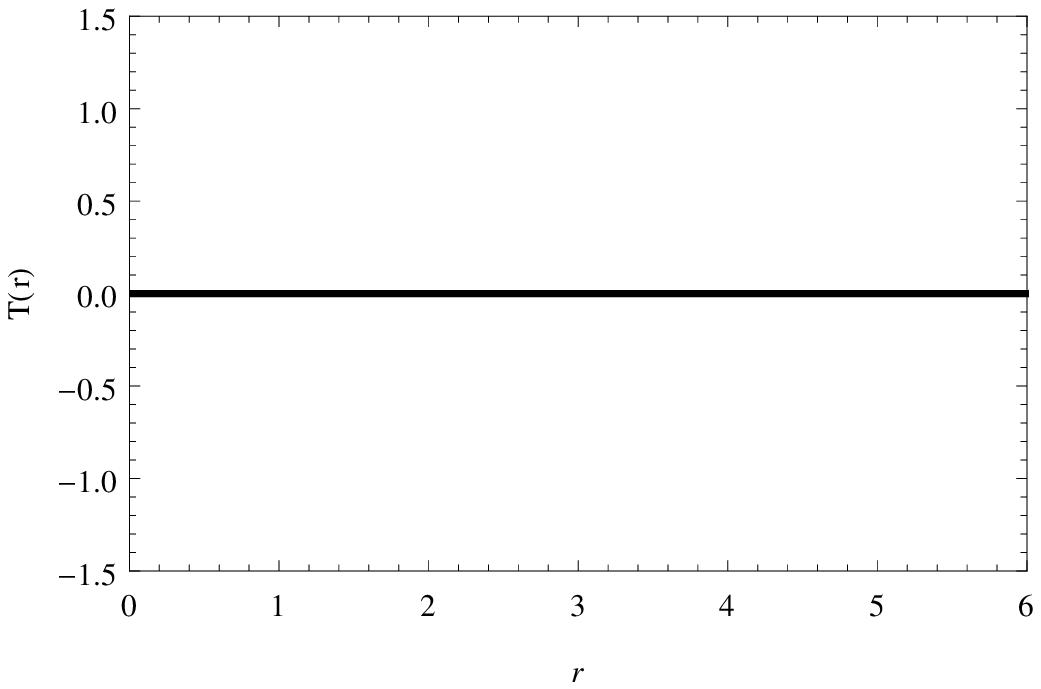}
\caption{{\it{The solutions for $G( r )^2$ of (\ref{G}), for the Ricci
scalar $R(r)$ of (\ref{Ricci}), for the Kretschmann scalar $R_{\mu \nu \rho
\sigma }R^{\mu \nu \rho \sigma }(r)$  of (\ref{RRRR}) and for the torsion
scalar $T(r)$ of (\ref{scalartorsion1}), as a function of $r$, for the
positive
(thick solid curve)  branch, for  $D=4$,
$\alpha=-1$, $Const =-1$ and $\Lambda=0$. Left graphs correspond
to charged solutions with $Q=1$, while right graphs correspond to
uncharged solutions with $Q=0$. The thin $0$-line is depicted for
convenience.}}}
\label{Lambdaeffnull}
\end{center}
\end{figure}

We consider $\Lambda=0$, which satisfies the condition $\Lambda_{eff}=0$
for the positive branch (as we mentioned below (\ref{Leff}) the
negative branch cannot lead to $\Lambda_{eff}=0$). 
In Fig.~\ref{Lambdaeffnull} from up to down we
show $G(r)^2$, the Ricci scalar, the Kretschmann scalar and the torsion
scalar $T(r)$ as a function of $r$, for the positive branch, with left
graphs corresponding to charged solutions ($Q=1$) and right graphs
corresponding to uncharged solutions ($Q=0$).

The charged solutions possess a physical singularity at $r=0$, where
all invariants diverge, however it is shielded a horizon at $r=r_H$, where
$G(r)^2$ becomes zero. Examining the Killing vector, and similarly to the
previous cases, we deduce that $r_H$ is a cosmological Killing horizon. 
In the uncharged case we observe that both the Ricci and torsion
scalars vanish, however the Kretschmann scalar diverges at $r=0$.  
Notice that in this case $G(r)^2$ has no roots, that is
there is not a horizon to shield the physical singularity at $r=0$, which
is therefore a naked one.

\end{itemize}

\section{Concluding  Remarks}
\label{conclusions}

In this work we considered D-dimensional $f(T)$ gravity including the 
Maxwell field. We extracted exact charged black-hole solutions depending
on the functional form of $f(T)$, on the electric charge and on the number
the dimensionality D. Finally, we investigated the singularities and the
horizons of the obtained solutions, following two different approaches.
Firstly, by studying the torsion invariants constructed using the 
Weitzenb{\"{o}}ck's connection and the vierbein solutions, and secondly by
studying the curvature invariants constructed using the Levi-Civita
connection and the metric solutions.

The main result is that in Maxwell-$f(T)$ gravity the curvature invariants
possess more physical singularities than the torsion ones, in some
particular solution subclasses. This difference disappears in the uncharged
case, or in the case where $f(T)$ gravity becomes the usual linear-in-$T$
teleparallel gravity, thus it reveals the novel behavior that is
introduced by the combined complication of the non-trivial $f(T)$ structure
with the electromagnetic sector. It seems that curvature and torsion
invariants behave very differently depending on the presence of the
Maxwell field. More generally, extending gravity in terms of $f(T)$ or
$f(R)$ formulations could give very different results as soon as matter
fields are taken into account.

Finally, we have to note that, in the scenario we have considered,  the
physical singularities are not always shielded by horizons. Thus, the
cosmic censorship does not always hold for D-dimensional Maxwell-$f(T)$
gravity. From a cosmological point of view such a feature could be
extremely relevant in order to investigate the early phases of cosmic
evolution. On the other hand, considering astrophysical structures in
strong field regimes, derived from torsion or curvature representation of
gravity, could give rise to very deep differences in dynamics
\cite{Boehmer004,stabile}.

 \vskip .2in \noindent {\large {\textbf{Acknowledgments}}}\\
S.C.  is supported by INFN (Sez. di Napoli). The research of E.N.S. is
implemented within the framework of the Action
Supporting Postdoctoral Researchers of the Operational Program
``Education and Lifelong Learning'' (Actions Beneficiary: General
Secretariat for Research and Technology), and is co-financed by
the European Social Fund (ESF) and the Greek State. Y. V. is supported by
FONDECYT grant 11121148, and by Direcci\'{o}n de Investigaci\'{o}n y
Desarrollo, Universidad de La Frontera, DIUFRO DI11-0071. \newline

\providecommand{\href}[2]{#2}

\begingroup

\raggedright

\endgroup

\end{document}